\documentclass[prd,twocolumn,nofootinbib,showpacs,superscriptaddress]{revtex4-1}

\usepackage{amsfonts}
\usepackage{amsmath}
\usepackage{amssymb}
\usepackage{bm}
\usepackage{dcolumn}
\usepackage{graphicx}
\usepackage{graphics}
\usepackage[latin1]{inputenc}
\usepackage{latexsym}
\usepackage{rotating}
\usepackage{hyperref}
\usepackage[all]{hypcap}
\usepackage{xspace} % Sensible space treatment at end of simple macros
\usepackage[usenames]{color}
\usepackage{ulem}
\definecolor {darkgreen}{rgb}{0.2,0.7,0.2}

\newcommand\be{\begin{equation}}
\newcommand\ba{\begin{eqnarray}}
\newcommand\ee{\end{equation}}
\newcommand\ea{\end{eqnarray}}

\newcommand\bw{\begin{widetext}}
\newcommand\ew{\end{widetext}}

\newcommand{\nn}{\nonumber}

\newcommand{\rot}{{\mbox{\tiny rot}}}

\newcommand{\ext}{{\mbox{\tiny ext}}}

\newcommand{\BL}{{\mbox{\tiny BL}}}

\newcommand{\rr}{{\mbox{\tiny rr}}}

\newcounter{subsubsubsection}[subsubsection]

\begin{document}
\title{Tidal heating and torquing of a Kerr black hole to
  next-to-leading order in the tidal coupling}

\author{Katerina Chatziioannou}
\affiliation{Department of Physics, Montana State University, Bozeman, Montana 59717, USA}

\author{Eric Poisson}
\affiliation{Department of Physics, University of Guelph, Guelph, Ontario, Canada NIG 2W1}

\author{Nicol\'as Yunes}
\affiliation{Department of Physics, Montana State University, Bozeman, Montana 59717, USA}

\date{\today}


% Other PACS
%04.20.Ex,  % Initial value problem, existence and uniqueness of
%solutions 
%0.4.20.Cv Fundamental problems and general formalism
% 04.70.Bw, % Classical black holes
 
% 97.60.Lf  % Black holes (Late stages of stellar evolution)

% Sometimes we want to include preprint numbers, let's put them here
\preprint{IGPG-05/11-3}

%-------------------------------------------------------------------------
\begin{abstract}
%-----------------------------------------------------------------------

We calculate the linear vacuum perturbations of a Kerr black hole
surrounded by a slowly varying external spacetime to third order in
the ratio of the black-hole mass to the radius of curvature of the
external spacetime. This expansion applies to two relevant physical
scenarios: (i) a small Kerr black hole immersed in the gravitational
field of a much larger external black hole, and (ii) a Kerr black hole
moving slowly around another external black hole of comparable
mass. This small-hole/slow-motion approximation allows us to
parametrize the perturbation through slowly varying, time-dependent
electric and magnetic tidal tensors, which then enable us to separate
the Teukolsky equation and compute the Newman-Penrose scalar
analytically to third order in our expansion parameter. We obtain
generic expressions for the mass and angular momentum flux through the
perturbed black hole horizon, as well as the rate of change of the
horizon surface area, in terms of certain invariants constructed from
the electric and magnetic tidal tensors. We conclude by applying these
results to the second scenario described above.  

%-----------------------------------------------------------------------
\end{abstract}
%-----------------------------------------------------------------------

\maketitle

%%%%%%%%%%%%%%%%%%%%%%%%%%%%%%%%%%%%%%%%%%%%%%%%%%%%%%%%%%%
\section{Introduction}
\label{intro}
%%%%%%%%%%%%%%%%%%%%%%%%%%%%%%%%%%%%%%%%%%%%%%%%%%%%%%%%%%%

The Kerr metric describes the external spacetime of a spinning black hole (BH) in general relativity and it is the most general, stationary and axisymmetric solution to the vacuum Einstein equations~\cite{Kerr:1963ud,Hawking:1971bv,Hawking:1971vc,PhysRev.164.1776,Israel:1967za,Carter:1971zc}. Real astrophysical BHs, however, are not in vacuum, but rather subject to vacuum, gravitational perturbations from surrounding objects. These perturbations will result in a change of the BH intrinsic parameters, characterized by mass and angular momentum flux through the horizon. Since this flux could affect the gravitational waves (GWs) emitted by BH binary systems through a modification to the balance law and, thus, the GW frequency evolution, it is important to fully understand the horizon dynamics involved.  

The effect of tidal perturbations on GWs emitted during BH inspirals varies depending on the system considered~\cite{Poisson:1994yf,Alvi:2001mx,Tagoshi:1997jy,PhysRevD.72.124016}. On the one hand, for quasicircular, comparable mass BH inspirals, the effect of these fluxes on the GW phase amounts to less than one radian for an event in the LIGO frequency band~\cite{Alvi:2001mx}. On the other hand, for quasicircular, extreme mass-ratio inspirals, such tidal effects can increase the duration of the signal up to 20 days in two years for an event in the frequency band of a space-borne detector~\cite{Hughes:2001jr,Martel:2003jj,Hughes:2001jr,Price:2001un}. For eccentric and inclined extreme mass ratio inspirals (EMRIs), this effect could be even larger~\cite{Pound:2007th,Yunes:2009ef,PhysRevD.83.044044}.

Poisson~\cite{Poisson:2004cw} developed a general formalism for calculating the linear, vacuum, gravitational perturbations of a BH owing to a dynamical external universe. In this formalism, one assumes the external universe is {\emph{slowly varying}}, such that the perturbations produced can be expanded in powers of the ratio of the BH mass to the radius of curvature of the external universe. This external universe is parametrized by the Weyl tensor, which can be written in terms of electric and magnetic tidal tensors and their derivatives. With these expansions, the equation that governs the evolution of linear perturbations, the Teukolsky equation, can be separated and perturbatively solved. Its solutions can then be used to compute the fluxes of mass and angular momentum across the perturbed horizon, as well as the increase in the horizon surface area. 

Poisson's work treats rotating and nonrotating BHs in two different frameworks: the {\emph{metric formalism}} (which is applicable to nonrotating BHs only) and the {\emph{curvature formalism}} (which is applicable to both rotating and nonrotating BHs). The metric formalism consists of directly perturbing the Schwarzschild metric and solving the linearized Einstein equations. This perturbed metric allows for the calculation of the Regge-Wheeler~\cite{Regge:1957rw} and Zerilli~\cite{Zerilli:1970la} master functions, which can then be used to calculate the horizon fluxes. The curvature formalism consists of perturbatively solving the time-domain Teukolsky equation~\cite{teukolsky,Press:1973zz,Teukolsky:1974yv} for the Newman-Penrose (NP) scalar $\psi_{0}$. This scalar is then used to obtain generic formulas for the horizon fluxes. Although the Schwarzschild case has been studied extensively mostly through the metric formalism~\cite{Poisson:2004cw,Poisson:2005pi,Martel:2005ir,Taylor:2008xy,Comeau:2009bz,Poisson:2009qj,Vega:2011ue}, the Kerr analysis through the curvature formalism has only been carried out to leading order in inverse powers of the radius of curvature of the external universe~\cite{Poisson:2004cw,Yunes:2005ve,Comeau:2009bz}.

The purpose of this paper is to solve for the linear vacuum perturbations and the associated tidal fluxes of a perturbed Kerr metric to next-to-leading order in inverse powers of the radius of curvature of the external universe using the curvature formalism. We use this formalism because a typical metric perturbation of the Kerr spacetime, in terms of tensor spherical harmonics, does not allow for the separation of the linearized Einstein equations in all coordinates. Instead, we choose to work with curvature perturbations, and more specifically with the NP scalar $\psi_0$, whose evolution equation, the Teukolsky equation, can be separated and thus perturbatively solved.

We begin by assuming that the external universe causes a vacuum
perturbation that is slowly varying. The Weyl tensor of the external
universe can then be decomposed into electric and magnetic tidal
tensors that depend only on time. After projecting this Weyl tensor
onto an appropriate tetrad, we obtain an asymptotic 
$\psi^{\rm asy}_{0}$ that automatically satisfies the Teukolsky equation in the 
asymptotic region. We construct a full $\psi_{0}$ by promoting the
radial dependence of $\psi^{\rm asy}_{0}$ to arbitrary functions of
radius. These functions are such that $\psi_{0}$ satisfies the
Teukolsky equation not just asymptotically, but down to the BH
horizon, while approaching $\psi^{\rm asy}_{0}$ in the asymptotic region. With
$\psi_0$ in hand, we can compute the mass and angular momentum fluxes,
as well as the increase in horizon area within the curvature
formalism.   

The main results of this paper are this fully analytic NP scalar, together with the mass and angular momentum fluxes derived from it in terms of certain invariants that parametrize the external universe geometry [see Eqs.~\eqref{Mdotffinal} and \eqref{Jdotffinal}]. These quantities scale as
\be
\langle \dot{M} \rangle  \sim {\cal{O}}(M^5/{\cal{R}}^5)\,, 
\ee
\be
\langle \dot{J} \rangle \sim {\cal{O}}(M^5/{\cal{R}}^4)+{\cal{O}}(M^6/{\cal{R}}^5)\,,
\ee
where ${\cal{R}}$ is the radius of curvature of the external spacetime, $M$ is the mass of the perturbed BH and $J$ the magnitude of its spin angular momentum. Similarly, the rate of change of
the horizon's surface area scales as [see Eq.~\eqref{Adotffinal}]
\be
\langle \dot{A} \rangle \sim {\cal{O}}(M^5/{\cal{R}}^4)+{\cal{O}}(M^6/{\cal{R}}^5)\,.
\ee

Although the Newman-Penrose scalar obtained here is valid for arbitrarily fast-rotating
backgrounds, certain assumptions used when computing the fluxes 
in the small-hole/slow-motion approximation prohibit us from taking the slow-rotation limit 
and comparing our results to Schwarzschild BH ones. To achieve this comparison, we modify 
these assumptions to obtain expressions for the fluxes valid in the slow-rotation limit. 
We show that the fluxes obtained with these new assumptions reduce correctly to the
ones obtained in the metric formalism when considering linear vacuum
perturbations of the Schwarzschild metric. We thus provide two sets of expressions 
for the tidal fluxes: one set valid for slowly rotating BHs and one set valid for
rapidly rotating BHs. The construction of a set of expressions valid 
uniformly for all rotations is left for future work.  
 
This paper is divided as follows: 
Sec.~\ref{pert-scheme} describes the curvature formalism and the perturbative scheme employed; 
Sec.~\ref{asymptotic} computes the asymptotic form of the Weyl tensor and the NP scalar;
Sec.~\ref{psi0} computes the exact form of the NP scalar $\psi_{0}$; Sec.~\ref{Psi-flux} computes the Teukolsky potential $\Psi$ and the mass and angular momentum fluxes; 
Sec.~\ref{slow-rot} computes the mass and angular momentum fluxes in the limit of a slowly rotating BHs; 
Sec.~\ref{binary} applies our results to certain astrophysically motivated scenarios; 
and Sec.~\ref{conclusions} concludes and points to future research. 

Henceforth, we employ the following conventions. We use geometrized units, where $G=1$ and $c=1$. 
The symbol ${\cal{O}}(a)$ stands for terms of relative order $a$. Greek indices range over spacetime coordinates, 
while Latin indices in between parentheses denote tetrad components. The Einstein summation convention is assumed all
throughout the paper, where repeated indices are to be summed over unless otherwise specified.

%%%%%%%%%%%%%%%%%%%%%%%%%%%%%%%%%%%%%%%%%%%%%%%%%%%%%%%%%%%
\section{The Curvature Formalism and The Small-Hole/Slow-Motion Approximation}
\label{pert-scheme}
%%%%%%%%%%%%%%%%%%%%%%%%%%%%%%%%%%%%%%%%%%%%%%%%%%%%%%%%%%%

\subsection{The curvature formalism}

The curvature formalism is described in detail in~\cite{Poisson:2004cw}, and here we present only the basic ideas that are relevant to this paper.

Let us consider a background Kerr BH with mass $M$ and spin angular momentum with magnitude $J=aM$. Its perturbations can be characterized by the NP scalar $\psi_0$, which can be written in terms of the perturbed Weyl tensor $\delta C_{\alpha \beta \gamma \delta}$ as
\be\label{psi-def}
\psi_{0} = - \delta C_{\alpha \beta \gamma \delta} l^{\alpha} m^{\beta} l^{\gamma} m^{\delta}\,,
\ee
where $l^{\alpha}$ and $m^{\alpha}$ are two tetrad $4$-vectors. This scalar, when computed in the Kinnersley tetrad, diverges at the unperturbed horizon. Therefore, it is more convenient to express $\psi_0$ in terms of the Hartle-Hawking tetrad. Using ingoing Kerr coordinates $(v,r,\theta,\psi)$, that are related to Boyer-Lindquist coordinates $(t_{\BL},r_{\BL},\theta_{\BL},\phi_{\BL})$ through the transformation
\begin{align}
v&=t_{\BL}+ \int \frac{r^2+a^2}{\Delta} dr \,, \qquad r=r_{\BL} \,, \\
\theta&= \theta_{\BL} \,, \qquad \psi=\phi_{\BL} + a \int \frac{1}{\Delta} dr \,,
\end{align}
the curvature variable $\Psi$ (the so-called {\emph{Teukolsky potential}}) is then free of divergences on the horizon \cite{Poisson:2004cw} and defined by
\be\label{Psi-def}
\Psi(v,\theta,\psi) 
= -\psi_0(\mbox{HH})\big|_{r=r_+} = 
-\frac{\Delta^2}{4(r^2+a^2)^2}\psi_0(\mbox{K}) \bigg|_{r=r_{+}}\,.
\ee
Here, $\psi_0(\rm K)$ and $\psi_{0}(\rm HH)$ are the NP scalars in the Kinnersley and Hartle-Hawking tetrad respectively, $\Delta=r^2-2Mr+a^2$ and $r_{\pm}=M\pm \sqrt{M^2-a^2}$ is the radial location of the unperturbed Kerr horizon. The axial symmetry of the Kerr solution allows us to decompose $\Psi$ in azimuthal modes
\begin{align}
\Psi(v, \theta, \psi) = \sum_m \Psi ^m (v, \theta) e^{i m \psi} \,,
\end{align}
that evolve independently.

Defining the \emph{integrated curvatures} 
\ba
\label{Phimplus}
\Phi^m_+(v,\theta) &=& e^{\kappa v} \int_{v}^{\infty} e^{-(\kappa - i m \Omega_H) v'} \Psi^m(v',\theta) dv'\label{phi+} \,, 
\\
\Phi^m_-(v,\theta) &=&  \int^{v}_{-\infty} e^{ i m \Omega_H v'} \Psi^m(v',\theta) dv' \label{phi-}\,,
\ea
where $\kappa = (r_+ - M)/(r^2_+ + a^2)$ is the surface gravity of the unperturbed Kerr horizon and $\Omega_H=a/(r^2_+ + a^2)$ is the angular velocity of the unperturbed horizon, and using the dynamics of the perturbed horizon, we can calculate the change in its surface area via~\cite{Poisson:2004cw}
\be\label{Adot-def}
\frac{\kappa}{8 \pi} \langle \dot{A} \rangle = \frac{r^2_+ + a^2}{2} \sum _{m}  \int \langle |\Phi^m_+|^2 \rangle \sin{\theta} d\theta \,.
\ee
Using the first law of BH mechanics 
\be
\frac{\kappa}{8 \pi} \langle \dot{A} \rangle = \langle \dot{M} \rangle - \Omega_{H}\langle \dot{J} \rangle\,,
\ee
and the fact that
\be
\langle \dot{M} \rangle_{m,\omega} =  \frac{\omega}{m}\langle \dot{J} \rangle_{m,\omega} \,,
\ee 
for any $(m,\omega)$ Fourier mode, the mass and angular momentum fluxes are \cite{Poisson:2004cw}
\begin{align}\label{Mdot-def}
\langle \dot{M} \rangle &= \frac{r^2_+ + a^2}{4 \kappa} \sum _{m } \left[ 2 \kappa \int \langle |\Phi^m_+|^2 \rangle \sin{\theta} d\theta \right. \nn \\
&- \left. i m \Omega_H \int \langle \bar{\Phi}^m_+ \Phi^m_- - \Phi^m_+ \bar{\Phi}^m_- \rangle \sin{\theta} d\theta \right] \,,
\end{align}
and
\begin{align}\label{Jdot-def}
\langle \dot{J} \rangle = - \frac{r^2_+ + a^2}{4 \kappa} \sum _{m \neq 0} (i m) \int \langle \bar{\Phi}^m_+ \Phi^m_- - \Phi^m_+ \bar{\Phi}^m_- \rangle \sin{\theta} d\theta \,.
\end{align}
%

%------------------------------------------------------------------------
\subsection{The small-hole/slow-motion approximation}

For a non rotating BH, there are only two relevant length scales: the mass of the BH $M$ and the radius of curvature of the external universe ${\cal{R}}$. The assumption that the external universe is varying slowly implies that
\be\label{appr-sch}
\frac{M}{{\cal{R}}} \ll 1\,.
\ee 
If we assume that our perturbed BH is in a circular orbit around some other object of mass $M_{\ext}$, then ${\cal{R}}$ is given by the GW wavelength, which leads to 
\be\label{appr-bin}
\frac{M}{{\cal{R}}} \sim \frac{M}{M + M_{\ext}} V^3 \ll 1 \,,
\ee
where $V$ is the orbital velocity.

The above requirement can be met in two ways: the {{\emph{small-hole approximation}} or the \emph{slow-motion approximation}}. In the former, $M/M_{\ext} \ll V^{-3}$, so our perturbed BH is immersed in the gravitational field of a much larger BH and the orbital velocity is unrestricted. Of course, since $V$ cannot exceed one, this condition translates to $M/M_{\ext} \ll 1$. In the slow-motion approximation, $V \ll (1+M_{\rm ext}/M)^{1/3}$ so the BH moves slowly and the masses of the BHs are unrestricted. Since the masses are positive, this condition can be met if one requires that $V \ll 1$. 

The rotating case is slightly different, since now we have two length scales associated with the perturbed BH. One of them is again $M$, while the second is the time scale associated with rotation
\be
T_{\rot} = \frac{1} {\Omega_{H}} = \frac{r_+^2 + a^2}{a} = \frac{ 2 M (1 + \sqrt{1 - \chi^2})}{\chi}\,,
\label{Trot-def}
\ee
where we have defined the dimensionless spin parameter $\chi=a/M$. The
requirement that the external universe be slowly varying implies that
both $M$ and $T_{\rot}$ be small compared to ${\cal{R}}$. The second
condition gives 
\be\label{appr-kerr}
\frac{M}{{\cal{R}}} \ll \chi\,,
\ee
and since $0 < \chi < 1$ for Kerr black holes, this is a stronger
constraint than Eq.~\eqref{appr-sch}. This, in turn, changes the
defining conditions of the slow-motion/small-hole approximations into 
\be
\frac{M}{M + M_{\ext}} V^3 \ll \chi \,.
\ee
Thus, in the small-hole approximation, we must require that $M/M_{\rm ext} \ll \chi/V^{3}$, while in the slow-motion approximation we must have $V \ll \chi^{1/3} (1+M_{\rm ext}/M)^{1/3}$. For equal-mass, slowly rotating BHs in a circular orbit, which come into contact with a velocity of roughly $V \sim 0.4$, the above condition puts a limit on how small $\chi$ can be, namely $\chi \gg 0.03$.

The above constraint prohibits us from taking the $\chi=0$ limit and comparing our results to those obtained within Schwarzschild BH perturbation theory~\cite{Poisson:2004cw}. The slowly rotating case requires a different analysis, where our expansion parameter no longer is forced to satisfy $M/{\cal{R}} \ll \chi$, but rather $M/{\cal{R}} = {\cal{O}}(\chi)$. Physically, this is because in the $\chi \to 0$ limit, the rotational time scale formally diverges [see Eq.~\eqref{Trot-def}], and thus, we cannot require that the radius of curvature of the external universe be much larger than it. The slow-rotation limit of the small-hole/slow-motion approximation is described in more detail in Sec.~\ref{slow-rot}. 

%%%%%%%%%%%%%%%%%%%%%%%%%%%%%%%%%%%%%%%%%%%%%%%%%%%%%%%%%%%
\section{Asymptotic Form of the Newman-Penrose Scalar}
\label{asymptotic}
%%%%%%%%%%%%%%%%%%%%%%%%%%%%%%%%%%%%%%%%%%%%%%%%%%%%%%%%%%%

Since the perturbations we are dealing with are of vacuum type, in the sense that there is no matter in the vicinity of the background BH, the effect of the external universe is fully described by its Weyl tensor in the neighborhood of the background BH. Our first task is, therefore, to determine its asymptotic form through order ${\cal{O}}({\cal{R}}^{-3})$.  

To perform this computation we may ignore the gravitational field of
the BH, which at $r \gg r_+$ is negligible compared to the
tidal effects of the external gravitational field. If we imagine that
the BH moves on a timelike geodesic $\gamma$ in the external
gravitational field, we may express the external Weyl tensor through a
Taylor expansion in powers of $r$ at fixed $v$~\cite{Poisson:2009qj}. The Taylor expansion is
implemented formally as   
\begin{equation} 
C_{\alpha\beta\gamma\delta} \!=\! g_{\alpha}^{\ \alpha'}\!
g_{\beta}^{\ \beta'}\! g_{\gamma}^{\ \gamma'}\! g_{\delta}^{\ \delta'}\!  
\left[\! C_{\alpha'\beta'\gamma'\delta'} \!
- \!C_{\alpha'\beta'\gamma'\delta';\epsilon'} \!\sigma^{\epsilon'} \!\!\!
+ \! O\!\!\left(\!\frac{r^2}{{\cal R}^4}\!\right) \!\right]\!\!, 
\end{equation} 
where $C_{\alpha\beta\gamma\delta}$ is the Weyl tensor evaluated at
a point $x$ off the worldline, while $C_{\alpha'\beta'\gamma'\delta'}$
is the Weyl tensor evaluated at the advanced point $x'$ on the world
line, which is linked to $x$ by a future-directed (from $x$ to $x'$)
null geodesic. The quantity $\sigma(x,x')$ is half the squared geodesic distance
between $x$ and $x'$, $\sigma_{\alpha'}$ is its gradient with respect
to $x^{\alpha'}$, and $g_{\alpha}^{\ \alpha'}(x,x')$ is the parallel
propagator from $x$ to $x'$. 

Following the methods described in Appendix D of~\cite{Poisson:2009qj}, 
we erect a tetrad $(u^{\alpha'}, e^{\alpha'}_a)$ on the worldline 
$\gamma$, which we parallel transport off the worldline with 
$g^{\alpha}_{\ \alpha'}$. Tensors on and off the worldline can be
decomposed on the tetrad, which we indicate, for example, via 
$C_{a0b0}(x) = C_{\alpha\gamma\beta\delta} 
e^\alpha_a u^\gamma e^\beta_b u^\delta$, in which 
$e^\alpha_a = g^{\alpha}_{\ \alpha'} e^{\alpha'}_a$ and 
$u^\alpha = g^{\alpha}_{\ \alpha'} u^{\alpha'}$. It may be shown that
the frame components of the parallel propagator differ from the
Minkowski metric by terms of order $r^2$ and higher, so that they play
no role in an expansion of the Weyl tensor through ${\cal{O}}(r)$. Making
use of $\sigma^{\alpha'} = r u^{\alpha'} - x^a e^{\alpha'}_a$ ---
Eq.~(D4) of ~\cite{Poisson:2009qj} --- we find that the frame
components of the Weyl tensor are 
\begin{subequations} 
\label{weyl_expansion}
\begin{align} 
C_{a0b0}(v,r,\theta,\psi) &= C_{a0b0}(v,0) - r \dot{C}_{a0b0}(v,0) \nn \\ 
&+ x^e C_{a0b0|e}(v,0) + O(r^2/{\cal R}^4), \\ 
C_{abc0}(v,r,\theta,\psi) &= C_{abc0}(v,0) - r \dot{C}_{abc0}(v,0) \nn \\
&+ x^e C_{abc0|e}(v,0) + O(r^2/{\cal R}^4), \\ 
C_{abcd}(v,r,\theta,\psi) &= C_{abcd}(v,0) - r \dot{C}_{abcd}(v,0) \nn \\
&+ x^e C_{abcd|e}(v,0) + O(r^2/{\cal R}^4), 
\end{align} 
\end{subequations} 
in which $x^a = [r\sin\theta\cos\psi, r\sin\theta\sin\psi,
r\cos\theta]$ is a Cartesian system defined in the usual way from the
spherical polar coordinates $(r,\theta,\psi)$, $\dot{C}_{a0b0} = 
C_{\alpha'\gamma'\beta'\delta';\epsilon'} e^{\alpha'}_a u^{\gamma'} 
e^{\beta'}_b  u^{\delta'} u^{\epsilon'}$, and $C_{a0b0|e} = 
C_{\alpha'\gamma'\beta'\delta';\epsilon'} e^{\alpha'}_a u^{\gamma'} 
e^{\beta'}_b  u^{\delta'} e^{\epsilon'}_e$, with the other projections
defined in an analogous way.

The frame components of the Weyl tensor and its derivatives evaluated
at $r=0$ can be expressed in terms of the irreducible components
provided by the tidal quadrupole moments ${\cal E}_{ab}$ and 
${\cal B}_{ab}$ and tidal octupole moments  
${\cal E}_{abc}$ and ${\cal B}_{abc}$, which are all
symmetric-tracefree tensors in the Cartesian system $x^a$. The
relevant relations are listed in Appendix D of ~\cite{Poisson:2009qj}, but
we repeat them here for completeness: 
\begin{subequations} \label{EB2}
\begin{align} 
C_{a0b0}(v,0) &= {\cal E}_{ab}, \\ 
C_{abc0}(v,0) &= \epsilon_{abp} {\cal B}^p_{\ c}, \\ 
C_{abcd}(v,0) &= -\epsilon_{abp} \epsilon_{cdq} {\cal E}^{pq}
\\
C_{a0b0|e}(v,0) &= {\cal E}_{ab|e}, \\ 
C_{abc0|e}(v,0) &= \epsilon_{abp} {\cal B}^p_{\ c|e}, \\ 
C_{abcd|e}(v,0) &= -\epsilon_{abp} \epsilon_{cdq} {\cal E}^{pq}_{\ \ |e}
\end{align} 
\end{subequations}  
with the covariant derivatives of the quadrupole tidal tensors
\begin{subequations}   \label{EB3}
\begin{align}
{\cal E}_{ab|c} = {\cal E}_{abc} + \frac{1}{3} \bigl( \epsilon_{acp}
\dot{\cal B}^p_{\ b} + \epsilon_{bcp} \dot{\cal B}^p_{\ a} \bigr), \\ 
{\cal B}_{ab|c} = \frac{4}{3} {\cal B}_{abc} - \frac{1}{3} \bigl( \epsilon_{acp}
\dot{\cal E}^p_{\ b} + \epsilon_{bcp} \dot{\cal E}^p_{\ a} \bigr)\,. 
\end{align}
\end{subequations}  
It is a straightforward task to insert the irreducible decompositions
within Eq.~(\ref{weyl_expansion}) to obtain an expansion of the Weyl
tensor through order $r$. 

We can now compute the NP scalar 
\begin{equation} 
\psi_0 = -C_{\alpha\gamma\beta\delta} l^\alpha m^\gamma
l^\beta m^\delta, 
\label{psi0_def}
\end{equation} 
where $l^\alpha$ and $m^\alpha$ are members of the null tetrad
involved in the decomposition of the Weyl tensor. In our computation,
the null tetrad at $(v,r,\theta,\psi)$ differs from the null tetrad at
$(v,0,\theta,\psi)$ by terms of ${\cal{O}}(r^2)$, which can be ignored. Therefore, 
the null vectors are given by $l^0 = 1$, $l^a = n^a = x^a/r$, and 
\begin{equation} 
m^a = \frac{1}{\sqrt{2}} (\cos\theta\cos\psi-i\sin\psi, 
\cos\theta\sin\psi + i\cos\psi, -\sin\theta). 
\end{equation} 
With this, we find that 
\be
\psi_0 = -C_{a0b0} m^a m^b  + 2 C_{abc0} n^a m^b m^c - C_{acbd} n^a m^c n^b m^d\,.
\ee

An explicit computation of $\psi_0$ through ${\cal{O}}(r)$ is facilitated
by the introduction of the quantities
\allowdisplaybreaks 
\begin{subequations} 
\label{alphabeta2} 
\begin{align} 
\alpha_{2,0} &= {\cal E}_{11} + {\cal E}_{22}, \\ 
\alpha_{2,\pm 1} &= {\cal E}_{13} \mp i {\cal E}_{23}, \\
\alpha_{2,\pm 2} &= {\cal E}_{11} - {\cal E}_{22} 
\mp 2i {\cal E}_{12}, \\   
\beta_{2,0} &= {\cal B}_{11} + {\cal B}_{22}, \\ 
\beta_{2,\pm 1} &= {\cal B}_{13} \mp i {\cal B}_{23}, \\
\beta_{2,\pm 2} &= {\cal B}_{11} - {\cal B}_{22} 
\mp 2i {\cal B}_{12}
\end{align}
\end{subequations} 
and 
\begin{subequations} 
\label{alphabeta3} 
\begin{align} 
\alpha_{3,0} &= {\cal E}_{113} + {\cal E}_{223}, \\ 
\alpha_{3,\pm 1} &= {\cal E}_{111} + {\cal E}_{122} 
\mp i \bigl( {\cal E}_{112} + {\cal E}_{222} \bigr), \\ 
\alpha_{3,\pm 2} &= {\cal E}_{113} - {\cal E}_{223} 
\mp 2 i {\cal E}_{123}, \\ 
\alpha_{3,\pm 3} &= {\cal E}_{111} - 3 {\cal E}_{122} 
\mp i \bigl( 3 {\cal E}_{112} - {\cal E}_{222} \bigr), \\
\frac{3}{4} \beta_{3,0} &= {\cal B}_{113} + {\cal B}_{223}, \\ 
\frac{3}{4} \beta_{3,\pm 1} &= {\cal B}_{111} + {\cal B}_{122} 
\mp i \bigl( {\cal B}_{112} + {\cal B}_{222} \bigr), \\ 
\frac{3}{4} \beta_{3,\pm 2} &= {\cal B}_{113} - {\cal B}_{223} 
\mp 2 i {\cal B}_{123}, \\ 
\frac{3}{4} \beta_{3,\pm 3} &= {\cal B}_{111} - 3 {\cal B}_{122} 
\mp i \bigl( 3 {\cal B}_{112} - {\cal B}_{222} \bigr),
\end{align}
\end{subequations} 
which act as substitutes for the linearly independent components of
the tidal moments. This calculation is also aided by the introduction of
the spin-weight $s=+2$ (scalar) spherical harmonics  
\begin{subequations} 
\label{Y2m} 
\begin{align} 
\mbox{}_2 Y_2^0 &= -\frac{3}{2} \sin^2\theta, \\ 
\mbox{}_2 Y_2^{\pm 1} &= -\sin\theta(\cos\theta \mp 1) 
e^{\pm i\psi}, \\  
\mbox{}_3 Y_2^{\pm 2} &= \frac{1}{4} (\cos\theta \mp 1)^2 
e^{\pm 2i\psi}, 
\end{align}
\end{subequations} 
and 
\begin{subequations} 
\label{Y3m} 
\begin{align} 
\mbox{}_2 Y_3^0 &= -\frac{5}{2} \sin^2\theta\cos\theta, \\ 
\mbox{}_2 Y_3^{\pm 1} &= \frac{5}{8} \sin\theta 
(3\cos\theta \pm 1)(\cos\theta \mp 1) e^{\pm i\psi}, \\ 
\mbox{}_2 Y_3^{\pm 2} &= \frac{1}{4} (3\cos\theta \pm 2) 
(\cos\theta \mp 1)^2 e^{\pm 2i\psi}, \\ 
\mbox{}_2 Y_3^{\pm 3} &= \frac{1}{8} \sin\theta 
(\cos\theta \mp 1)^2 e^{\pm 3i\psi}. 
\end{align}
\end{subequations} 
These functions are mutually orthogonal, but for our convenience they are not normalized. 

Putting all of this together, we find that the asymptotic form of the Weyl scalar is
given by  
\begin{align} 
\psi_0 &\sim \psi^{\rm asy}_0 \equiv -\sum_m z_{2 m} 
\, \mbox{}_2 Y_2^m(\theta,\psi) 
- r \sum_m z_{3 m}
\, \mbox{}_2 Y_3^m(\theta,\psi) 
\nonumber \\ & \quad \mbox{} 
+ \frac{1}{3} r \sum_m \dot{z}_{2 m} \, \mbox{}_2 Y_2^m(\theta,\psi)
+ O(r^2/{\cal R}^4)\,,
\label{asy-psi0} 
\end{align}
where we have defined
\be\label{z-general}
z_{\ell m}(v) \equiv \alpha_{\ell m}(v) + i \beta_{\ell m}(v)\,, 
\ee
and an overdot indicates differentiation with respect to $v$. 
The asymptotic NP scalar satisfies the Teukolsky equation in the
asymptotic region, which we have verified explicitly.  

An alternative way to derive $\psi^{\rm asy}_{0}$ goes as follows. One
starts with the metric of a vacuum spacetime expressed in null 
coordinates $(v,r,\theta,\phi)$ in a neighborhood of a geodesic worldline 
situated at $r=0$; this metric is presented in Sec.~III B of
Ref.~\cite{Poisson:2009qj}. One then computes the Weyl tensor for this
metric, and projects it onto the tetrad described previously. The
end result is again Eq.~\eqref{asy-psi0}. We have carried out the
calculation both ways and found agreement.

%%%%%%%%%%%%%%%%%%%%%%%%%%%%%%%%%%%%%%%%%%%%%%%%%%%%%%%%%%%
\section{The Full Newman-Penrose Scalar}
\label{psi0}
%%%%%%%%%%%%%%%%%%%%%%%%%%%%%%%%%%%%%%%%%%%%%%%%%%%%%%%%%%%

In this section, we compute the full NP scalar $\psi_{0}$ associated with a perturbed spinning BH metric. 
It is defined by Eq.~\eqref{psi-def} and in what follows, we use its asymptotic form constructed in Sec~\ref{asymptotic} to determine the full scalar that solves the Teukolsky equation everywhere inside $r \ll {\cal{R}}$, including at $r=r_+$. 

\subsection{Teukolsky equation} 

It is convenient to work with the Fourier transform of $\psi_0$, which can be resolved in mode functions that separate the variables:
\begin{align}\label{psi0-f-dec}
\tilde{\psi}_0 &=  \sum_{\ell m} \;  \tilde{z}_{\ell m}(\omega) R_{\ell m}(r) \; {}_{2}S^{\ell m}(\theta) e^{i m \psi}\,,
\end{align}
where $R_{\ell m}(r)$ are free functions of radius, ${}_{2}S^{\ell
  m}(\theta)$ are spin-weight $+2$ spheroidal harmonics and the
overhead tilde stands for the Fourier transform. The angular functions
reduce to spin-weight $s=+2$ (scalar) spherical harmonics in the limit
as $\psi_{0}$ becomes time independent (the zero-frequency
limit). Notice that all advanced time derivatives of $\psi_0$ simply
pull down factors of $-i \omega$ when working in terms of the Fourier
transform.  

This separation of variables allows us to decouple the Teukolsky equation in Kerr coordinates into an angular equation~\cite{Teukolsky:1972le,Teukolsky:1974yv}
\begin{align}
& \frac{1}{\sin{\theta}}  \partial_{\theta} \left(\sin{\theta} \; \partial_{\theta}{}_{2}S^{\ell m} \right) + 
\left( - \frac{m^{2}}{\sin^{2}{\theta}} - 4 a \omega \cos{\theta} 
\right. 
\nn \\
&- \left. \frac{4 m \cos{\theta}}{\sin^{2}{\theta}} - 4 \cot^{2}{\theta} + E^{\ell m} - 4 \right) \!\! \;{}_{2}S^{\ell m} = {\cal{O}}(\omega^{2}), 
\label{ang-eq}
\end{align}
and a radial equation~\cite{Teukolsky:1972le,Teukolsky:1974yv}
\begin{align}
& \Delta \; \partial_{rr} R^{\ell m} + 2 \left[3 (r - M) - i K\right] \partial_{r} R^{\ell m}
\nn \\
&+ \left[
- \frac{8 i (r - M) K}{\Delta} + 6 i \omega r - \lambda \right] R^{\ell m} = {\cal{O}}(\omega^{2})\,.
\label{eq:Radial-EOM}
\end{align}
Here, $E^{\ell m}$ is a separation constant, we have defined 
$K \equiv \left(r^{2} + a^{2}\right) \omega - a m$, $\lambda \equiv E^{\ell m} - 2 a m \omega - 6$, 
linearized in $\omega$ and we have set $s = +2$. These expressions correct a typo in Eq.~$(2.10)$ of~\cite{Teukolsky:1974yv}: the term explicitly proportional to $\omega$ inside the square brackets of Eq.~\eqref{eq:Radial-EOM}
should really be $2 (2s-1) i \omega r$, instead of $-2 (2s+1)i \omega r$ in Eq.~$(2.10)$ of~\cite{Teukolsky:1974yv}.

One can recast the radial equation in a more amenable form by transforming to
\be
x \equiv \frac{r - r_{+}}{r_{+} - r_{-}}\,.
\label{reduced-radial-coord}
\ee
Doing so, the radial equation becomes
\allowdisplaybreaks
\begin{widetext}
\begin{align}
& x (x + 1) \partial_{xx}R^{\ell m} +  \left(6x + 3 + 2 i \gamma m\right)\partial_{x} R^{\ell m}+ \left[4i \gamma m\frac{(2x + 1)}{x (1 + x)}  - (\ell+3)(\ell-2) \right] R^{\ell m}\nn \\
& \quad \mbox{}
- 2i\omega (r_+ - r_-) \biggl\{ 
\bigl[ x(1+x) + (2\rho-1)(x+\rho) \bigr]\partial_{x} R^{\ell m} 
+ \biggl[ \frac{2(2x+1) \bigl[ x(1+x) + (2\rho-1)(x+\rho)
  \bigr]}{x(1+x)} 
\nonumber \\ & \quad \mbox{}
- 3(x+\rho) 
-im\gamma \frac{\ell(\ell+1) + 4}{\ell(\ell+1)} \biggr] R^{\ell m} \biggr\}={\cal{O}}(\omega^2)\,. 
 \label{radialeq}
\end{align}
\end{widetext}
where we have defined $\gamma \equiv a/(r_{+} - r_{-})$, $\rho \equiv r_{+}/(r_{+} - r_{-})$ and we have used the eigenvalue 
\be
E^{\ell m} = \ell \left(\ell + 1\right) - \frac{8 a m}{\ell(\ell+1)} \omega + {\cal{O}}(\omega^{2})\,,
\label{Eeq}
\ee
in the $M \omega \ll 1$ limit for $s = +2$. Note that this equation reduces exactly to Eq.~$(9.25)$ in~\cite{Poisson:2004cw} when $\omega = 0$ and $\ell =2$.

The rest of this section is devoted to finding the solutions to the angular and radial sectors of the Teukolsky equation to ${\cal{O}}(\omega)$. Notice though that up until now no assumption has been made on the rate of rotation of the background black hole, i.e.~$\chi$ is unrestricted and we have only expanded in $M \omega \ll 1$.  

%----------------------------------------
\subsection{Angular sector}
Spin-weighted spheroidal harmonics satisfy the angular equation. In general, they must be solved for numerically, but in the limit of small $\omega$ one can represent them as a finite sum of spin-weighted spherical harmonics of the same weight~\cite{Press:1973zz}
\begin{align}
\;{}_{2}S^{\ell m} &= \; {}_{2}Y^{\ell m}(\theta,\psi) -  4 a \omega \sum_{\ell \neq \ell'} \sqrt{\frac{2 \ell + 1}{2 \ell' + 1}}
\nn \\
&\times  \frac{\left< \ell 1 m 0|\ell' m\right> \left<\ell 1 \,-2 0| \ell' \, -2\right>}{\ell (\ell+1) - \ell' (\ell' + 1)} {}_{2}Y^{\ell' m}(\theta,\psi) \,,
\end{align}
whose eigenvalues were given in Eq.~\eqref{Eeq}.  The quantity $\left<j_{1} j_{2} m_{1} m_{2} | J M\right>$ are the usual Clebsch-Gordan coefficients. 

In this paper, we are interested in the $\ell = 2$ and $\ell = 3$
modes only. With our choice of normalization for the spin-weighted
spherical harmonics, the angular functions are given by
\begin{align}
_{2}S^{20} &= \;{}_{2}Y^{20} - \frac{2}{5} a \omega \;{}_{2}Y^{30}\,,
\\ 
_{2}S^{2 \pm 1} &= \left(\;{}_{2}Y^{2\pm 1} + \frac{16}{45} a \omega \;{}_{2}Y^{3 \pm 1} \right) e^{\mp i \psi}\,, 
\\
_{2}S^{2 \pm 2} &= \left(\;{}_{2}Y^{2\pm 2} - \frac{2}{9} a \omega \;{}_{2}Y^{3 \pm 2} \right) e^{\mp 2 i \psi}\,, 
\\
_{2}S^{30} &= \;{}_{2}Y^{30}
% - \frac{4}{63} a \omega^{30} \;{}_{2}Y^{20} - \frac{1}{14} a \omega^{30} \;{}_{2}Y^{40}\,,
\\
_{2}S^{3\pm 1} &= \;{}_{2}Y^{3\pm 1}
%- \frac{16}{63} a \omega^{3 \pm 1} \;{}_{2}Y^{2\pm 1} - \frac{3}{28} a \omega^{3 \pm 1} \;{}_{2}Y^{4\pm 1} 
\; e^{\mp i \psi}\,,
\\
_{2}S^{3\pm 2} &= \;{}_{2}Y^{3\pm 2}
% + \frac{40}{63} a \omega^{3 \pm 2} \;{}_{2}Y^{2\pm 2} - \frac{3}{2} a \omega^{3 \pm 2} \;{}_{2}Y^{4\pm 2}
\; e^{\mp 2 i \psi}\,,
\\
_{2}S^{3\pm 3} &= \;{}_{2}Y^{3\pm 3}  
%- \frac{1}{4} a \omega^{3 \pm 3} \;{}_{2}Y^{4\pm 3} 
\; e^{\mp 3 i \psi}\,.
\end{align}
The $\ell=3$ modes do not require any ${\cal{O}}(\omega)$ corrections. For future convenience, we will also write the spheroidal harmonics as 
\be
{}_{2}S^{\ell m} = ({}_{2}Y^{\ell m} + a \omega \; b^{\ell m} \; {}_{2}Y^{(\ell+1)m}) e^{\mp i m \psi}\,, 
\label{eq:decomp-Sm}
\ee
where 
\begin{align}
\left(b^{20},b^{2\pm1},b^{2\pm2}\right) &= \left(-\frac{2}{5},\frac{16}{45},-\frac{2}{9}\right), \quad b^{3m}=0 \,.
\label{b2m-coeffs}
\end{align}
We have checked that these spheroidal harmonics satisfy the angular equation in Eq.~\eqref{ang-eq} with the eigenvalues of Eq.~\eqref{Eeq} to ${\cal{O}}(\omega^2)$.

%----------------------------------------
\subsection{Radial sector}
\label{subsec:RadialSector}

Usually, when working to all orders in $\omega$, Eq.~\eqref{radialeq} is solved numerically, after imposing certain boundary conditions at spatial infinity and at the horizon. When working to linear order in $\omega$, however, solutions can be found analytically in terms of hypergeometric functions~\cite{Press:1973zz,Mano:1996vt,Mino:1997bw}. When expressed in closed form, these solutions are rather complicated and require the imposition of certain recursion relations. For this reason, we will rederive them here and present explicit solutions at ${\cal{O}}(1)$ and ${\cal{O}}(\omega)$ for the $\ell = 2$ and $\ell =3$ modes separately. 

\subsubsection{Expansion of the radial function} 

Let us then decompose the radial functions as 
\be
R^{\ell m} = R^{\ell m,0} + \omega R^{\ell m,1} 
+ {\cal O}(\omega^2) \,,
\label{Radial-decomp}
\ee
where $R^{\ell m,0}$ and $R^{\ell m,1}$ are assumed to be $\omega$ independent. With this decomposition, the radial equation to ${\cal{O}}(\omega^{0})$ becomes
\begin{align}
& \left\{ x (x + 1) \partial_{xx} + \left[3 \left(2x + 1\right) + 2 i \gamma m
\right] \partial_{x} \right.
\nn \\
&+ \left. \frac{4}{x (1 + x)} \left[
i \gamma m \left(1  + 2x \right)
\right. \right. 
\nn \\
&- \left. \left.
\frac{1}{4} (\ell+3)(\ell-2)  x \left(x + 1\right)
\right]\right\}R^{\ell m,0} = 0\,.
\label{0th-order-eq}
\end{align}
Note that this equation reduces exactly to Eq.~$(9.25)$ in~\cite{Poisson:2004cw} when $\ell = 2$. Similarly, the radial equation to linear order in $\omega$ is
\begin{align}
& \Big\{ x (x + 1) \partial_{xx} + \left[3 \left(2x + 1\right) + 2 i \gamma m
\right] \partial_{x} \Big.
\nn \\
&+ \left. \frac{4}{x (1 + x)} \left[
i \gamma m \left(1  + 2x \right)
\right. \right. 
\nn \\
&- \left. \left.
\frac{1}{4} (\ell+3)(\ell-2)  x \left(x + 1\right)
\right]\right\}R^{\ell m,1} = T^{\ell m}\,,
\label{1st-order-eq}
\end{align}
where the source $T^{\ell m}$ has nothing to do with a matter source, but rather it is constructed from the zeroth-order solution $R^{\ell m,0}$ and its first time derivative, namely

\begin{align}
T^{\ell m} &\equiv 2i (r_+ - r_-) \biggl\{ 
\bigl[ x(1+x) + (2\rho-1)(x+\rho) \bigr]\partial_{x} R^{\ell m,0} \nn \\
&+ \biggl[ \frac{2(2x+1) \bigl[ x(1+x) + (2\rho-1)(x+\rho)
  \bigr]}{x(1+x)} 
\nonumber \\ & \quad \mbox{}
- 3(x+\rho) 
-im\gamma \frac{\ell(\ell+1) + 4}{\ell(\ell+1)} \biggr] R^{\ell m ,0} \biggr\}\,. 
\end{align}

%

%----------------------------------------
\subsubsection{Boundary conditions}

Asymptotically, we know that $\psi_{0}$ should approach Eq.~\eqref{asy-psi0} as $r \gg r_{+}$. Let us then impose this condition to see what the boundary conditions on $R_{\ell m}$ should be at spatial infinity. Fourier transforming Eq.~\eqref{asy-psi0} and setting this equal to Eq.~\eqref{psi0-f-dec} we have
\begin{align}\label{psi_0as}
&\sum_{m} \left[\tilde{z}_{2m} \;{}_{2}Y^{2m} + \frac{i \omega r}{3} \tilde{z}_{2m} \;{}_{2}Y^{2m} + \tilde{z}_{3m} r \;{}_{2}Y^{3m}\right] \sim \nn \\
& \sum_{m} \!\left[ \tilde{z}_{2m} R_{2m}\!\left({}_{2}Y^{2m} \!+\!
    b^{2m} a \omega \,{}_{2}Y^{3m} \right) \!+\! \tilde{z}_{3m} R_{3m}
  \,{}_{2}Y^{3m} \right]\,,
\end{align}
where we have replaced $\tilde{\dot{z}}_{2m} = -i \omega \tilde{z}_{2m}$ and we have rewritten the spheroidal harmonics
as in Eq.~\eqref{eq:decomp-Sm}.  

Let us now look at different $\ell$ modes separately. Clearly, all terms proportional to ${}_{2}Y^{2m}$ must be set to zero independently from those proportional to ${}_{2}Y^{3m}$ . Let us then decompose the radial function as in Eq.~\eqref{Radial-decomp} with $\ell = 2$, so as to separate the term of ${\cal{O}}(1)$ from the term of ${\cal{O}}(\omega)$. For the $(\ell,m)=(2,m)$ mode to ${\cal{O}}(1)$, we have
\be
R^{2m,0} \sim 1\,,
\label{R2m0-BC}
\ee
while to ${\cal{O}}(\omega)$ we find
\be
R^{2m,1} \sim \frac{i}{3} r\,.
\label{R2m1-BC}
\ee
Similarly, for the $(\ell,m)=(3,m)$ mode we have
\be
R^{3m,0} \sim r\,,
\label{R3m0-BC}
\ee
to leading order in $\omega$. 

The radial function must also satisfy a regularity condition at the
event horizon. Because $R^{\ell m}(r)$ is associated with a Weyl
scalar $\psi_0$ constructed with the Kinnersley tetrad, which is
singular on the event horizon, the radial function itself cannot be
expected to be nonsingular at $r=r_+$. Instead, the relation of
Eq.~(\ref{Psi-def}) implies that it is $\Delta^2 R^{\ell m}(r)$ that
must be smooth at $r = r_+$. In terms of the new radial coordinate
$x$, this means that $x^2 R^{\ell m}(x)$ must be smooth at $x=0$.  

%----------------------------------------
\subsubsection{Radial solution: $m=0$ case}

Let us first concentrate on the solution when $m=0$, as this is easier than when $m \neq 0$. To leading order in $\omega$, we find the solutions
\begin{align}
R^{20,0} &= C^{20,0}_{1} + C^{20,0}_{2} \left[ 6 \ln \left(\frac{x}{1+x} \right) 
\right. 
\nn \\
&+ \left.
\frac{(2x+1) (6x^{2}+6 x - 1)}{2 x^{2}(1+x)^{2}}\right] \,
\label{eq:R20-sol-homg}
\\
R^{30,0} &= C^{30,0}_{1} \left(2 x + 1 \right) + C^{30,0}_{2} \left[ 60 \left(2 x + 1\right) \ln\left(\frac{x}{1+x}\right)
\right. 
\nn \\
&+ \left. \frac{240 x^{3} + 130 x^{2} + 120 x^{4} + 10x - 1}{x^{2}(1+x)^{2}}  \right]\,,
\label{eq:R30-sol-homg}
\end{align}
where $C^{20,0}_{1,2}$ and $C^{30,0}_{1,2}$ are constants of integration. 

Let us now impose the boundary conditions in Eqs.~\eqref{R2m0-BC} and~\eqref{R3m0-BC}. Asymptotically expanding the two solutions, about $x = \infty$, we find
\begin{align}
R^{20,0} &\sim C^{20,0}_{1} + {\cal{O}}(x^{-5})\,,
\\
R^{30,0} &\sim 2 C^{30,0}_{1} x + {\cal{O}}(1)\,.
\end{align}
Imposing Eqs.~\eqref{R2m0-BC} and~\eqref{R3m0-BC},
\be
C^{20,0}_{1} = 1\,, \qquad
C^{30,0}_1 = \frac{1}{2}(r_+-r_-) = \frac{a}{2\gamma}\,. 
\ee

The selection of $C^{20}_2$ and $C^{30}_2$ is based on the requirement
that $x^2 R^{\ell m}(x)$ be smooth at $x = 0$. Expanding the
solutions about $x=0$, we find that 
\begin{align}
R^{20,0} &\sim C^{20,0}_{2} \left[ -\frac{1}{2 x^{2}} + \frac{3}{x} + 6 \ln{x} + {\cal{O}}(1) \right]\,,
\\
R^{30,0} &\sim C^{30,0}_{2} \left[- \frac{1}{x^{2}} + \frac{12}{x} + 60 \ln{x} + {\cal{O}}(1)
\right]\,.
\end{align}
To eliminate the logarithmic terms, which would break smoothness of the Teukolsky function
at the event horizon, we must choose $C^{20,0}_{2} = 0 = C^{30,0}_{2}$. To leading order in $\omega$, the
solutions then become 
\be
R^{20,0} = 1\,
\qquad
R^{30,0} = \frac{a}{\gamma} \left(x + \frac{1}{2}\right)\,.
\ee

Let us now move on to the $\ell=2$ mode of the radial solution to linear order in $\omega$. The ${\cal{O}}(\omega)$ piece of the $\ell=3$ solution is not needed as it would contribute to ${\cal{O}}({\cal{R}}^{-4})$ in $\psi_0$. The $\ell=2$ solution contains a homogeneous piece plus an inhomogeneous piece. The homogeneous piece is the same as that of Eq.~\eqref{eq:R20-sol-homg}. Including the inhomogeneous solution, we find
\begin{align}
R^{20,1} \!\!&= C^{20,1}_{1}\!\! + \! C^{20,1}_{2} \! \!\left[6 \ln\!\!\left(\!\frac{x}{1 + x}\!\right) \!\!
+\!\!\frac{(2x + 1)(6x^{2}+6x-1)}{2 x^{2}(1 + x)^{2}} \right]
\nn \\
&+ \ln(x) \left(\!\frac{i a}{\gamma} + \!4 i \gamma a + 2 i M \right)
- \ln(x+1) \left(\frac{ia}{\gamma} + 4 i \gamma a \right)
\nn \\
&+ \frac{i a}{3 \gamma} x - i M (1+x)^{-1} - \frac{i M}{6} (1 + x)^{-2}\,,
\end{align}
where $C^{20,1}_{1,2}$ are constants of integration.

Let us now impose the boundary conditions. Asymptotically expanding about $x = \infty$, we find
\begin{align}
R^{20,1} &\sim \frac{i a}{3 \gamma} x + 2 i M \ln(x) + C^{20,1}_{1} + {\cal{O}}[x^{-1}] \nn \\
&\sim \frac{i}{3} r  - \frac{i}{3}  r_{+} + 2 i M \ln(x) + C^{20,1}_{1}\,.
\end{align}
The solution automatically satisfies the boundary conditions in Eq.~\eqref{R3m0-BC}. We are free to choose $C^{20,1}_{1}$ as we wish, since it only affects the subleading behavior of our solution. Here we choose to leave it arbitrary, since it will not affect the final solution, as will become clear in the analysis of Sec.~\ref{Psi-flux}.

Asymptotically expanding the $\ell=2$ solution about $x = 0$, we find
\be
R^{20,1}\! \sim \!-\! \frac{C^{20,1}_{2}}{2 x^{2}} + \frac{3 C^{20,1}_{2}}{x} + \ln(x)\! \left(\!6 C^{20,1}_{2}\! +\! 2 i M \!+\! \frac{i a}{\gamma} \!+\! 4 i \gamma a\!\right).
\ee
The above solution possesses two types of divergences at the horizon: a logarithmic and a polynomial one. The requirement that the Teukolsky potential be smooth at the event horizon forces us to eliminate the logarithmic one by setting
\be
C^{20,1}_{2} = -\frac{2 i \gamma a}{3} - \frac{i a}{6 \gamma} - \frac{i M}{3}\,.
\ee
%

%----------------------------------------
\subsubsection{Radial solution: $m\neq0$ case}

Let us first work to ${\cal{O}}(1)$. The solutions to Eq.~\eqref{0th-order-eq} are
\be
R^{\ell m,0} = R^{\ell m,0}_{h,1} + R^{\ell m,0}_{h,2}\,,
\ee
where we have defined the two sets of homogeneous solutions
[$(R^{2m,0}_{h,1},R^{2m,0}_{h,2})$ for the $\ell=2$ mode and
$(R^{3m,0}_{h,1},R^{3m,0}_{h,2})$ for the $\ell=3$ mode] via
\begin{align}
R^{2m,0}_{h,1} &= A^{2m}_{h,1} x^{-2} (1 + x)^{-2} F(-4,1; -1 + 2 i m \gamma; -x)\,, 
\label{homg-sol-l=2}
\\
R_{h,2}^{2m,0} &= A^{2m}_{h,2}\left( 1 + \frac{1}{x} \right)^{2 i \gamma m}\label{homg2-sol-l=2}\,,
\\
R_{h,1}^{3m,0} &= A^{3m}_{h,1} x^{-2} (1 + x)^{-2} F(-5,2; -1 + 2 i m \gamma; -x)\,, 
\label{homg-sol-l=3}
\\
R_{h,2}^{3m,0} &=A^{3m}_{h,2}(2 m \gamma +3 i + 6 i x)  \left( 1 + \frac{1}{x} \right)^{2 i \gamma m}\,,
\end{align}
and $ F(a,b;c;z)$ is the hypergeometric function. 

Let us now impose smoothness and the appropriate boundary conditions. 
The homogeneous solutions $R_{h,2}^{2m,0}$ and $R_{h,2}^{3m,0}$ lead to a Teukolsky potential
that is not smooth at $x = 0$. We therefore set the integration constants $A^{2m}_{h,2}=0=A^{3m}_{h,2}$ 
such that these solutions do not contribute. Then, by setting
\begin{align}
A^{2m}_{h,1} &= - \frac{i}{6} m \gamma \left(1 + i m \gamma \right) \left(1 + 4 m^{2} \gamma^{2}\right)\,,
\label{A2m-const}
\\
A^{3m}_{h,1} &= -\frac{i}{180} m a \left(1 + i m \gamma \right) 
\left(1 + 4m^{2} \gamma^{2}\right) \left(3 + 2 i m \gamma\right)\,. 
\end{align}
$R^{2m,0}_{h,1}$ and $R^{3m,0}_{h,1}$ are normalized such that $R^{2m,0}_{h,1} \to 1$ and $R^{3m,0}_{h,1} \to r$ as $x \to \infty$. Moreover, these solutions have a well-defined Taylor expansion about $x = 0$. Note that no natural logarithm terms appear here, although clearly they both diverge as $x^{-2}$ as $x \to 0$. 

Let us now concentrate on the $\ell=2$ mode of the solution to ${\cal{O}}(\omega)$. As before, the $\ell=3$ mode would contribute  to $\psi_0$ at ${\cal{O}}({\cal{R}}^{-4})$, and we thus neglect it. The $\ell=2$ solution to ${\cal{O}}(\omega)$ will contain a homogeneous plus an inhomogeneous piece. The homogeneous one will be proportional to Eqs.~\eqref{homg-sol-l=2} and \eqref{homg2-sol-l=2}, while the inhomogeneous piece is the particular solution that can be obtained with the method of variation of constants by
\begin{align}\label{gen-sol}
R^{2m,1}_p &= - R_{h,1}^{2 m , 0}(x) \int \frac{R_{h,2}^{2 m , 0}(\xi) T^{2 m} (\xi)} {\xi (\xi+1) W(\xi)} d\xi \nn\\ 
&+ R_{h,2}^{2 m , 0}(x)  \int \frac{R_{h,1}^{2 m , 0}(\xi) T^{2 m} (\xi)} {\xi (\xi+1) W(\xi)} d\xi \,, 
\end{align}
where $W$ is the Wronskian associated with $R^{2m,0}_{h,1}$ and $R^{2m,0}_{h,2}$.
The first integrand in Eq.~\eqref{gen-sol} is a polynomial in $\xi$ and can be integrated immediately. 
The second term leads to a sum of hypergeometric functions, which can be manipulated with the identity
\begin{align} 
a(b\!-\!c) z\,&F(a+1, b ; c+1; z)
\!=\! c(c\!-\!1)\, F(a-1, b ; c-1; z) \nn \\
&- c \bigl[ c-1 + (a-b) z \bigr]\,  F(a, b ; c; z), 
\end{align}   
which can be derived from the Gauss relations for hypergeometric functions. 
By repeated application of this identity, the second term
can be expressed in a form that features 
$\mbox{}_2 F_1[1,1;3+2im\gamma;-x]$ only. 
The full inhomogeneous solution then has the form
\be
R^{2m,1}=A R^{2m,0}_{h,1} + B R_{h,2}^{2m,0} + R^{2m,1}_p \,,
\ee
where $A$ and $B$ are constants. As before, we set $B = 0$ by requiring smoothness of the Teukolsky function at the event horizon.  

The choice of $A$ deserves further discussion. In the limit as $x \to \infty$ the particular solution behaves as $R^{2m,1}_p \to  i r/3$, while the homogeneous solution $R^{2m,0}_{h,1} \to 1$. Therefore, the asymptotic value of $R^{2m,1}$ at infinity is specified by $R^{2m,1}_p$. The constant $A$ determines the subleading behavior of the solution at spatial infinity. What should we choose this subleading behavior to be? The answer to this question can be discovered by studying the $\chi \to 0$ limit. One can check that the particular solution diverges in this limit, but there is no physical reason for this to be the case. Indeed, the Teukolsky equation was solved under the assumption that $M \omega \ll 1$, and with $\omega = {\cal{O}}({\cal{R}}^{-1})$, this condition becomes $M/{\cal{R}}\ll1$. At no point did we expand in $\chi \ll 1$, and thus, there is no physical reason for the limit to be singular. Therefore, we will choose $A$ such that this singular behavior is eliminated:
\be\label{A-const}
A=\frac{i r_+}{6 A^{2m}_{h,1}}(2\rho-1) + i M \alpha\,,
\ee
where $A^{2m}_{h,1}$ is given by Eq.~\eqref{A2m-const} and $\alpha$ is an arbitrary constant, which in principle could depend on $m$ and $\chi$ through terms of ${\cal{O}}(\chi^{0})$ or higher. The factor multiplying it was inserted for later convenience. Notice that the first term in $A$ fixes the slow-rotation limit of $R^{2m,1}$, while the second accounts for extra freedom to modify the fast-rotation limit. 

With this choice of $A$, the inhomogeneous solution is
\begin{align}
\label{inhomg-sol}
R^{2m,1} & = \frac{ i M }{ 1 + \kappa_m }\frac{x^2}{(x+1)} F(1, 1; 3 + 2 i \gamma m;-x) 
\nn \\
&+ \frac{1}{(x+1)^{2}} \left[\frac{A}{x^{2}} + \frac{C_1}{x} + C_{2} + C_3 x + C_4 x^2 + C_{5} x^3  \right]\,, 
%&+ \frac{ i a}{3 \gamma } \frac{x^3 } {(x+1)^2} + C_4 \frac{x^2}{(x\!+\!1)^2}+C_3 \frac{x} {(x+1)^2 }
%\nn \\
%&+C_2 \frac{1}{ (x+1)^2} + C_1 \frac{1}{x (x+1)^2} +A \frac{1}{x^2(1+x)^2}\,, 
\end{align}
in which $\kappa_m = im\gamma$ and
\begin{widetext}
\begin{subequations} 
\begin{align} 
C_1 &= \frac{ (1+\kappa_m)(1+2\kappa_m)\kappa_m [ 10\kappa_m^3 - (5+18\rho) \kappa_m + 3\rho(8\rho-1)]}{9i\rho(1-2\kappa_m)} \;r_{+} + A^{2m}_{h,1}\frac{4}{2\kappa_m-1}\,, \\ 
C_2 &= \frac{128\kappa_m^4 \!+\! (164-264\rho) \kappa_m^3\!+\! (4-552\rho+576\rho^2) \kappa_m^2 \!-\! (41+294\rho-720\rho^2) \kappa_m\!-\! (9-30\rho-72\rho^2) }{36 i \rho(1-2\kappa_m) }\;r_{+} \nn\\
&+i M \alpha (1+\kappa_m) (1+2\kappa_m) \,,\\ 
C_3 &= \frac{16\kappa_m^4 + (20-24\rho) \kappa_m^3 - 24 \rho (3-4\rho) \kappa_m^2 - (5+42\rho-96\rho^2) \kappa_m- (1-10\rho+8\rho^2)}{2 i \rho (1+2\kappa_m)(1-2\kappa_m)}\;r_{+} + i M \alpha (1+\kappa_m)  \,,\\ 
C_4 &= \frac{ 32\kappa_m^4 + 32 \kappa_m^3+ (4-120\rho+192\rho^2) \kappa_m^2 - (8+72\rho-144\rho^2) \kappa_m 
- 3(1-2\rho)(1-4\rho)  }{6 i \rho(1+\kappa_m)(1+2\kappa_m)(1-2\kappa_m) } \;r_{+} + i M \alpha\,, \\
C_{5} &=\frac{ i a}{3 \gamma }\,.
\end{align} 
\end{subequations}
\end{widetext} 
Since the hypergeometric function behaves
as $\ln(x)/x$ when $x \gg 1$, in the limit $x \to \infty$ the above solution behaves as $R^{2m,1} \to  i r/3$ and satisfies the boundary condition in Eq.~\eqref{R3m0-BC}.

%---------------------------------------------------------------
\subsection{Summary of results}
The Fourier transform of the full NP scalar can be written as
\begin{align}\label{psi_0}
\tilde{\psi}_0 &= -\sum_{m} \tilde{z}_{2m} R_{h,1}^{2m,0}\;{}_{2}Y^{2m} - \sum_{m} \tilde{z}_{3m} R_{h,1}^{3m,0} \;{}_{2}Y^{3m}  \nn \\
&- \sum_{m} \omega \tilde{z}_{2m} \left(R^{2m,1} \;{}_{2}Y^{2m} + a \, b^{2m} R_{h,1}^{2m,0} \;{}_{2}Y^{3m} \right)\,,
\end{align}
where the radial functions are given by Eqs.~\eqref{homg-sol-l=2}, \eqref{homg-sol-l=3}, \eqref{inhomg-sol}
and the $b^{2m}$ coefficients are given by Eq.~\eqref{b2m-coeffs}, with the (unnormalized) spin-weighted spherical harmonics of Eqs.~\eqref{Y2m} and \eqref{Y3m}. One can easily compute $\psi_{0}$ by inverse Fourier transforming the above equation.

%%%%%%%%%%%%%%%%%%%%%%%%%%%%%%%%%%%%%%%%%%%%%%%%%%%%%%%%%%%
\section{The Teukolsky Potential \\ and the Fluxes}
\label{Psi-flux}
%%%%%%%%%%%%%%%%%%%%%%%%%%%%%%%%%%%%%%%%%%%%%%%%%%%%%%%%%%%

%-----------------------------------------------
\subsection{The Teukolsky potential}

The Teukolsky potential can be calculated from Eq.~\eqref{Psi-def}. In terms of the quantity $x$, defined in Eq. \eqref{reduced-radial-coord}, the Fourier-transformed potential $\tilde{\Psi}$ is given by
\be
\tilde{\Psi}(\omega,\theta^A)=-\frac{ x^2 (1+x)^2}{4 [ ( x  + \rho)^2 + \gamma^2  ]^2} \tilde{\psi}_0(\rm K)\bigg|_{x=0} \,.
\label{eq:Teuk-Pot-FT}
\ee
To return to the time domain and express $\Psi(v,\theta^{A})$ as a function of $v$, we can inverse Fourier-transform $\tilde{\Psi}(\omega,\theta^{A})$. One can see that $\Psi(v, \theta^{A})$ is given by $\tilde{\Psi}(\omega,\theta^{A})$ with the replacement $\omega \tilde{z}_{2m} \to i \dot{z}_{2m}$. Decomposing $\Psi(v,\theta^{A})$ in Fourier modes
\begin{align}
\Psi(v,\theta, \psi) = \sum_m \Psi ^m (v, \theta) e^{i m \psi} \,.
\end{align}
and evaluating the right-hand side of Eq.~\eqref{eq:Teuk-Pot-FT}, we obtain
\begin{align}\label{psi-m}
\Psi^m(v,\theta)&=\frac{ \sigma^4}{(1+\sigma)^2} ({z}_{2m} -  \alpha M \dot{z}_{2m} )A^{2m}_{h,1} \;{}_{2}Y^{2m}(\theta,0) \nn \\
&- \frac{1}{6} \frac{ \sigma^3}{1+\sigma} M \dot{z}_{2m} \;{}_{2}Y^{2m}(\theta,0) \nn\\
&+ i \frac{ \sigma^4}{(1+\sigma)^2}\chi M A^{2m}_{h,1}\dot{z}_{2m} b^{2m} \;{}_{2}Y^{3m}(\theta,0)\nn\\
&+\frac{1}{15}\frac{\sigma^5}{(1+\sigma)^2} M A^{2m}_{h,1} (3+2\kappa_m){z}_{3m} \;{}_{2}Y^{3m}(\theta,0)\,,
\end{align}
where $\sigma=\sqrt{1-\chi^2}$. The freedom to choose $\alpha$ is now seen to correspond to the freedom to redefine the tidal moments according to
\be
z_{2m} \to z_{2m} - \alpha M \dot{z}_{2m} + {\cal{O}}({\cal{R}}^{-4}).
\ee
With $\alpha$ complex and $m$ dependent, the freedom is complete: each component $z_{2m}$ can be redefined independently of any other components. In the following, we restrict this freedom by demanding that $\alpha$ be real and independent of $m$. In this case, the redefinition is the same for each component, and it corresponds to the uniform tensorial redefinitions
\begin{align}
{\cal{E}}_{ab} & \to {\cal{E}}_{ab} - \alpha M \dot{{\cal{E}}}_{ab} + {\cal{O}}({\cal{R}}^{-4}) \,, \\
{\cal{B}}_{ab} & \to {\cal{B}}_{ab} - \alpha M \dot{{\cal{B}}}_{ab} + {\cal{O}}({\cal{R}}^{-4}) \,, 
\end{align}
of the tidal moments.

The integrals in Eqs.~\eqref{phi+} and~\eqref{phi-} can be evaluated perturbatively to the appropriate order. Assuming that $\Psi^m$ varies slowly on a time scale $\tau = (\dot{\Psi}^m/\Psi^m)^{-1} \sim \cal{R}$ (since the time dependence of $\Psi$ comes from external universe perturbations), we can write for the $m \neq 0$ case,
\be\label{phi+-series}
\Phi^m_+\! =\! \frac{e^{i m \Omega_H} \Psi^m}{\kappa \!-\! i m \Omega_H}\! \left[1 \!+\!  \frac{1}{(\!\kappa \!-\! i m \Omega_H\!) \tau} \!+\! {\cal{O}}\left(\!\frac{1}{(\!\kappa \!-\! i m \Omega_H\!)^2 \tau^2}\!\right)\right],
\ee
\be\label{phi--series}
\Phi^m_- = \frac{ e^{i m \Omega_H} \Psi^m}{ i m \Omega_H} \left[1 +  \frac{i}{( m \Omega_H) \tau} + {\cal{O}}\left(\frac{1}{( m \Omega_H)^2 \tau^2}\right)\right].
\ee
The assumption under which the above expressions were derived, namely that $\Psi^m$ varies slowly, makes use of the stronger condition given by Eq.~\eqref{appr-kerr}. Therefore, the slow rotation limit is inaccessible in all results that follow. This limit requires an alternative treatment, which will be presented in Sec.~\ref{slow-rot}.

Let us discuss the expressions for $\Phi^{m}_{\pm}$ further. The function outside the square brackets has terms of ${\cal{O}}(M/{\cal{R}}^2)$ (proportional to $z_{2m}$) and ${\cal{O}}(M^2/{\cal{R}}^3)$ (proportional to $z_{3m}$ and $\dot{z}_{2m}$), while the second term in the square brackets is of ${\cal{O}}(M/{\cal{R}})$. Therefore, combining this with the leading-order terms that scale as ${\cal{O}}(M/{\cal{R}}^2)$, one finds terms of ${\cal{O}}(M^2/{\cal{R}}^3)$. The next-order terms scale as ${\cal{O}}(M^2/{\cal{R}}^2)$ and the lowest-order corrections that these induce will be of ${\cal{O}}(M^3/{\cal{R}}^4)$, which can thus be neglected.

Substituting the expression for $\Psi^m$ in Eqs.~\eqref{phi+-series} and \eqref{phi--series}, expressing $\kappa$ and $\Omega_H$ in terms of $M$ and $a$, and using the dimensionless parameter $\chi$, we find
\begin{align}\label{phim+}
\Phi^m_+ &= \frac{2 e^{i m \Omega_H} M \sigma^3 \;{}_2Y^{2m}}{(1+\sigma)(i m \chi- \sigma)}\left\{ -\sigma z_{2m} A^{2m}_{h,1}+\frac{M \dot{z}_{2m}}{6(i m \chi- \sigma)}\right.\nn\\
&\left.\times [12\sigma(1+\sigma)A^{2m}_{h,1}+(i m \chi- \sigma)(1+\sigma +6\sigma \alpha A^{2m}_{h,1})]\right\}\,,
\\
\Phi^m_- &=-\frac{2 i e^{i m \Omega_H} M \sigma^3 \;{}_2Y^{2m}}{(1+\sigma)m \chi}\left\{\sigma z_{2m} A^{2m}_{h,1}+\frac{M \dot{z}_{2m}}{6 m \chi}\right.\nn\\
&\left.\times [12 i \sigma(1+\sigma)A^{2m}_{h,1}- m \chi(1+\sigma +6\sigma \alpha A^{2m}_{h,1})]\right\}\,,
\label{phim-}
\end{align}
plus terms that contribute to ${\cal{O}}(M^2/{\cal{R}}^{4})$. 

The canceling of subleading terms in the above equations must be done carefully, keeping in mind that $\Phi^m_{\pm}$ in the end has to be squared and averaged over. Initially, $\Phi^{m}_{\pm}$ have terms that go as
\begin{align}
\Phi^m_{\pm} &\sim z_{2m} \;{}_2Y^{2m} +\dot{z}_{2m} \;{}_2Y^{2m} + z_{3m} \;{}_2Y^{3m} + \dot{z}_{2m} \;{}_2Y^{3m} \nn \\
&+ \dot{z}_{2m} \;{}_2Y^{2m} +\ddot{z}_{2m} \;{}_2Y^{2m}+ \dot{z}_{3m} \;{}_2Y^{3m} + \ddot{z}_{2m} \;{}_2Y^{3m} \nn \\
&+ \ddot{z}_{2m} \;{}_2Y^{2m} + {\cal{O}} \left( M^3/ {\cal{R}}^5 \right)\,.  
\end{align}
The terms in the first line come from $\Psi^m$, the ones in the second line come from $\dot{\Psi}^m$ and those in the third line come from $\ddot{\Psi}^m$. When squaring the above quantity, the two lowest-order terms will come from the squaring of the $z_{2m} \;{}_2Y^{2m}$ term and the cross term between $z_{2m} \;{}_2Y^{2m}$ and $\dot{z}_{2m} \;{}_2Y^{2m}$. They will be of ${\cal{O}} (M^2/{\cal{R}}^4)$ and ${\cal{O}} (M^3/{\cal{R}}^5)$ respectively. All other terms can be neglected to the order of approximation considered here.

%--------------------------------
\subsection{The fluxes}

Substituting Eqs.~\eqref{phim+} and \eqref{phim-} into Eqs.~\eqref{Mdot-def} and \eqref{Jdot-def}, we find 
\begin{align}\label{Mdotfinal}
\langle \dot{M} \rangle = \frac{\chi M^5}{45}  \sum_{m \neq 0}\frac{\langle \Im( \bar{z}_{2m} \dot{z}_{2m} ) \rangle(\sigma^2+m^2\chi^2)(4\sigma^2+m^2\chi^2)}{m} \,,
\end{align}
and
\begin{align}\label{Jdotfinal}
\langle \dot{J} \rangle &= -\frac{ \chi M^5}{45}\sum_{m \neq 0}  (\sigma^2+m^2\chi^2)(4\sigma^2+m^2\chi^2) \langle |z_{2m}|^2\rangle 
\nn \\
&+ \frac{2M^6}{45 m} \sum _{m \neq 0} \left\{\langle \Re(\bar{z}_{2m} \dot{z}_{2m})\rangle m \chi [(\sigma^2+m^2\chi^2)
\right. \nn \\
&\left. \times(4\sigma^2+m^2\chi^2) \alpha-2\sigma (1+\sigma)(2\sigma^2+m^2\chi^2)]
\right. \nn \\
&+\left. \langle \Im(\bar{z}_{2m} \dot{z}_{2m})\rangle [(3m^4\chi^4+13m^2\chi^2\sigma^2-4\sigma^4)(1+\sigma)]\right\}\,.
\end{align}
Notice that this agrees exactly with Eq. (9.32) in ~\cite{Poisson:2004cw}. Finally, using Eq.~\eqref{Adot-def} or the first law of BH mechanics, we can calculate the change of the horizon area
\begin{align}\label{Adotfinal}
\langle \dot{A} \rangle &= \frac{ 8 \pi\chi^2 M^5}{45 \sigma}\sum_{m \neq 0}  (\sigma^2+m^2\chi^2)(4\sigma^2+m^2\chi^2) \langle |z_{2m}|^2\rangle 
\nn \\
&-\frac{16 \pi \chi M^6}{45 \sigma m} \sum _{m \neq 0}\{\langle \Re(\bar{z}_{2m} \dot{z}_{2m})\rangle m \chi [(\sigma^2+m^2\chi^2)\nn\\
&\times(4\sigma^2+m^2\chi^2) \alpha-2 \sigma(1+\sigma)(2\sigma^2+m^2\chi^2)]
\nn \\
&+ \langle \Im(\bar{z}_{2m} \dot{z}_{2m})\rangle[2(1+\sigma)(m^4\chi^4+4m^2\chi^2\sigma^2-4\sigma^4)]\} \,.
\end{align}

One might notice that the above results do not depend on the $\ell=3$ part of $\Psi^m$. This is because the fluxes are obtained by squaring $\Phi$. Any $\ell=3$ terms in $\Psi^m$ result in quantities that scale as ${\cal{O}}(M^7/{\cal{R}}^6)$ in $\langle \dot{J} \rangle$ and as ${\cal{O}}(M^6/{\cal{R}}^6)$ in $\langle \dot{M} \rangle$, and they are beyond the order of approximation considered here. 

We can rewrite these expressions using Eqs.~\eqref{alphabeta2}, \eqref{alphabeta3}, \eqref{z-general} and the invariants
\allowdisplaybreaks
\begin{align}
E_1 = {\cal{E}}_{ab}{\cal{E}}^{ab}  \,,  \quad \qquad &B_1 = {\cal{B}}_{ab}{\cal{B}}^{ab}  \,,\\
E_2 = {\cal{E}}_{ab} s^b {\cal{E}}^{a}_{c} s^c \,, \qquad &B_2 = {\cal{B}}_{ab} s^b {\cal{B}}^{a}_{c} s^c  \,,\\
E_3 = ({\cal{E}}_{ab} s^a s^b)^2  \,, \qquad  &B_3 = ({\cal{B}}_{ab} s^a s^b)^2  \,,\\
E_4 = \epsilon_{pqc}{\cal{E}}^{pa}\dot{{\cal{E}}}^{q}_{a} s^c \,, \qquad &B_4 = \epsilon_{pqc}{\cal{B}}^{pa}\dot{{\cal{B}}}^{q}_{a} s^c \,,\\
E_5 = \epsilon_{pqc}{\cal{E}}^{p}_{a}\dot{{\cal{E}}}^{q}_{b} s^a s^b s^c  \,, \quad &B_5 = \epsilon_{pqc}{\cal{B}}^{p}_{a}\dot{{\cal{B}}}^{q}_{b} s^a s^b s^c \,,
\end{align}
where $s^a=(0,0,1)$ is the direction of the BH spin. Doing so we find

\begin{align}\label{Mdotffinal}
\langle \dot{M} \rangle &= \langle \dot{M}^{(5)} \rangle\,, 
\end{align}
\be\label{Jdotffinal}
\langle \dot{J} \rangle=\langle \dot{J}^{(4)} \rangle+\langle \dot{J}^{(5)}_{\alpha} \rangle+\langle \dot{J}^{(5)} \rangle\,,
\ee
\be\label{Adotffinal}
\langle \dot{A} \rangle=\langle \dot{A}^{(4)} \rangle+\langle \dot{A}^{(5)}_{\alpha} \rangle+\langle \dot{A}^{(5)} \rangle\,,
\ee
where we have defined
\begin{widetext}
\begin{subequations}
\begin{align}
\langle \dot{M}^{(5)} \rangle &=\frac{2 M^5 \chi}{45}\left[ -4\left(3\chi^2 + 1 \right) \langle E_4 + B_4 \rangle + 15 \chi^2\langle E_5+ B_5 \rangle \right]\,,
\\
\langle \dot{J}^{(4)} \rangle &= - \frac{2 M^5 \chi}{45} [8 (1 + 3 \chi^2) \langle E_1 + B_1 \rangle - 3 (4 + 17 \chi^2) \langle E_2 + B_2 \rangle + 15 \chi^2 \langle E_3 + B_3 \rangle ]\,,
\\
\langle \dot{J}^{(5)}_{\alpha} \rangle&= \frac{2 M^6 \chi \alpha}{45} [8 (1 + 3 \chi^2) \langle \dot{E}_1 + \dot{B}_1 \rangle 
- 3 (4 + 17 \chi^2) \langle \dot{E}_2 + \dot{B}_2 \rangle + 15 \chi^2 \langle \dot{E}_3 + \dot{B}_3 \rangle ]\,,
\\
\langle \dot{J}^{(5)} \rangle &=- \frac{4 M^6(\sigma+ 1 ) }{45}[ 4 \sigma \chi(\chi^2+1) \langle \dot{E}_1 \!+\! \dot{B}_1 \rangle - 3 \sigma \chi ( 3\chi^2 + 2) \langle \dot{E}_2\! +\! \dot{B}_2 \rangle+ 3 \sigma\chi^3 \langle \dot{E}_3 \!+\! \dot{B}_3\rangle   \nn\\
&-4(1 - 15 \chi^2 +2\chi^4)\langle E_4 \!+\! B_4\rangle - 3 \chi^2 (13 + 2 \chi^2) \langle E_5 \!+\! B_5 \rangle ]\,,
\\
\langle \dot{A}^{(4)} \rangle&=  \frac{16 \pi M^5 \chi^2}{45 \sigma} [8 (1 + 3 \chi^2) \langle E_1 + B_1 \rangle - 3 (4 + 17 \chi^2) \langle E_2 + B_2 \rangle + 15 \chi^2 \langle E_3 + B_3 \rangle ]\,,
\\
\langle \dot{A}^{(5)}_{\alpha} \rangle&=-\frac{16 \pi M^6 \chi^2 \alpha}{45\sigma} [8 (1 + 3 \chi^2) \langle \dot{E}_1 + \dot{B}_1 \rangle- 3 (4 + 17 \chi^2) \langle \dot{E}_2 + \dot{B}_2 \rangle + 15 \chi^2 \langle \dot{E}_3 + \dot{B}_3 \rangle ]\,,
\\
\langle \dot{A}^{(5)} \rangle &= \frac{32 \pi M^6(\sigma+ 1 ) }{45 \sigma}[ 4 \sigma \chi^2(\chi^2+1) \langle \dot{E}_1 \!+\! \dot{B}_1 \rangle - 3 \sigma \chi^2 ( 3\chi^2 + 2) \langle \dot{E}_2\! +\! \dot{B}_2 \rangle+ 3 \sigma\chi^4 \langle \dot{E}_3 \!+\! \dot{B}_3\rangle  \nn\\
& -8\chi(1 - 6 \chi^2 +\chi^4)\langle E_4 \!+\! B_4\rangle - 6 \chi^3 (4 +  \chi^2) \langle E_5 \!+\! B_5 \rangle ]\,.
\end{align}
\end{subequations}
\end{widetext}
In the above expressions, the superscript denotes the order of each term in $1/{\cal{R}}$. For example, $\langle \dot{M}^{(5)} \rangle$ is of ${\cal{O}}(M^5/{\cal{R}}^5)$ and $\langle \dot{J}^{(4)} \rangle $ is of ${\cal{O}}(M^5/{\cal{R}}^4)$. The subscript $\alpha$ represents terms that depend on the integration constant $\alpha$. The presence of these terms ($\langle \dot{J}^{(5)}_{\alpha} \rangle$ and $\langle \dot{A}^{(5)}_{\alpha} \rangle$) means that the equations are subject to an ambiguity related to the choice of $\alpha$. To resolve the ambiguity, we need to perform a matching calculation between the perturbed metric associated with the $\psi_{0}$ computed here and a PN metric valid far from the BH. This procedure determines $\alpha$ along with the explicit expressions for the tidal moments in terms of the post-Newtonian specification of the orbit~\cite{Alvi:1999cw,Yunes:2005nn,Yunes:2006iw,Taylor:2008xy,Gallouin:2012kb}.

%--------------------------
\subsection{Rigid-rotation limit} 
\label{rigid}
When the tidal field is in a state of rigid rotation with angular
frequency $\Omega$, the Weyl scalar $\psi_0$ depends on $\psi$ and $v$
through the combination $\psi - \Omega v$ only. This implies that the
tidal moments ${\cal E}_{ab}$ and ${\cal B}_{ab}$ satisfy the
relations 
\begin{equation} 
z_{2m}(v) = z_{2m}^\sharp e^{-i m \Omega v} \,, 
\end{equation} 
in which $z_{2m}^\sharp$ are constant amplitudes. It is then easy to
verify that with tidal moments of this form, the invariants 
satisfy the identities
\begin{subequations}\label{rigid-relations} 
\begin{align} 
\dot{E}_1&=\dot{E}_2=\dot{E}_3=0\,,\\
E_4 &= \Omega (2 E_1 - 3 E_2 )\,, \\  
E_5 &= \Omega ( E_2 -  E_3 )\,. 
\end{align}
\end{subequations} 
Similar identities are satisfied by the magnetic tidal tensor invariants. Therefore, in the case of rigid rotation, the terms containing $\alpha$ are identically zero $\langle \dot{J}^{(5)}_{\alpha} \rangle_{\rr}=\langle \dot{A}^{(5)}_{\alpha} \rangle_{\rr}=0$, and one is left with
\begin{widetext}
\begin{subequations}
\begin{align}
\langle \dot{M}^{(5)} \rangle_{\rr} &=-\frac{2 M^5 \Omega \chi}{45}\left[8 (1 + 3 \chi^2) \langle E_1 + B_1 \rangle - 3 (4 + 17 \chi^2) \langle E_2 + B_2 \rangle + 15 \chi^2 \langle E_3 + B_3 \rangle\right]\,,
\\
\langle \dot{J}^{(4)} \rangle_{\rr} &= - \frac{2 M^5 \chi}{45} [8 (1 + 3 \chi^2) \langle E_1 + B_1 \rangle - 3 (4 + 17 \chi^2) \langle E_2 + B_2 \rangle + 15 \chi^2 \langle E_3 + B_3 \rangle ]\,,
\\
\langle \dot{J}^{(5)} \rangle_{\rr} &= \frac{4 M^6(\sigma+ 1 )\Omega }{45}[8(1 - 15 \chi^2 +2\chi^4)\langle E_1 \!+\! B_1\rangle -3(4 - 73 \chi^2 +6\chi^4)\langle E_2 \!+\! B_2\rangle - 3 \chi^2 (13 + 2 \chi^2) \langle E_3 \!+\! B_3 \rangle ]\,,
\\
\langle \dot{A}^{(4)} \rangle_{\rr}&=  \frac{16 \pi M^5 \chi^2}{45 \sigma} [8 (1 + 3 \chi^2) \langle E_1 + B_1 \rangle - 3 (4 + 17 \chi^2) \langle E_2 + B_2 \rangle + 15 \chi^2 \langle E_3 + B_3 \rangle ]\,,
\\
\langle \dot{A}^{(5)} \rangle_{\rr} &= -\frac{32 \pi M^6(\sigma+ 1 )\Omega }{45 \sigma}[16\chi(1 - 6 \chi^2 +\chi^4)\langle E_1 \!+\! B_1\rangle-6\chi(4 - 28 \chi^2 +3\chi^4)\langle E_2 \!+\! B_2\rangle - 6 \chi^3 (4 +  \chi^2) \langle E_3 \!+\! B_3 \rangle ]\,.
\end{align}
\end{subequations}
\end{widetext}
Clearly then, when considering the realistic astrophysical scenario of a quasicircular binary black hole inspiral in the slow-motion approximation, the fluxes will not suffer from any ambiguity in $\alpha$. We present results for this particular scenario in Sec.~\ref{binary}.

%%%%%%%%%%%%%%%%%%%%%%%%%%%%%%%%%%%%%%%%%%%%%%%%%%%%%%%%%%%%%5
\section{The Slow-Rotation Limit}
\label{slow-rot} 

Although the $\psi_{0}$ calculated here is perfectly well behaved in the $\chi \to 0$ limit, 
the integrated curvatures, and therefore the fluxes, make use of the small-hole/slow-motion approximation,
which makes this limit inaccessible [see e.g.~Eq.~\eqref{appr-kerr}]. In particular, this implies that
Eqs.~\eqref{Mdotffinal} and \eqref{Jdotffinal} will not reduce to Eqs. (8.38) and (8.39) of
~\cite{Poisson:2004cw} in the $\chi \to 0$ limit. 

The mathematical reasons for why this limit is inaccessible become clearer if one considers the
assumptions involved in the evaluation of Eqs.~\eqref{phi+} and
\eqref{phi-}. As discussed in Sec.~\ref{Psi-flux}, we can approximate
those integrals with Eqs.~\eqref{phi+-series} and \eqref{phi--series}
by assuming that the exponentials vary rapidly compared to $\Psi^m$,
which is justified if $M/{\cal{R}} \ll \chi$. Then, each integration
by parts brings out a term that is ${\cal{O}}({\cal{R}}^{-1})$ smaller
than the previous one. This method, however, breaks down when the
exponentials vary slowly compared to $\Psi^m$, as in the $\Omega_H
\sim \chi \to 0$ limit, which forces Eq.~\eqref{phi--series} to
diverge.  

The above discussion suggests that the small $\chi$ limit of the integrated potentials and fluxes requires a special treatment. 
In this limit, the time scale associated with $\Omega_{\rm H} \simeq \chi/(4M)$ is comparable to ${\cal R}$: 
$\chi = {\cal{O}}(M/{\cal {R}})$. The horizon fluxes can be computed as an expansion in powers of $\chi$ and treating $M\omega = {\cal{O}}(M/{\cal {R}})$ as a quantity of first order in $\chi$. In this section, we calculate specifically the rate of change of the BH's surface area, which can be given by a simple expression; expressions for $\langle\dot{M}\rangle$ and $\langle\dot{J}\rangle$ are more complicated, and must be written in terms of integrals of the tidal tensors. 

Using $\chi$ instead of $\omega$ as an expansion parameter and keeping in mind that $a\omega = O(\chi^2)$, the small $\chi$ limit of Eq.~\eqref{psi-m} is
\begin{align} 
\Psi^m &= -\frac{1}{12} M \bigl( i m \Omega_{\rm H} z_{2m} 
+ \dot{z}_{2m} \bigr)\, \mbox{}_2 Y_2^m(\theta,0)+ O(\chi^2) \,,
\end{align} 
in the time domain.

The calculation of $\langle\dot{A}\rangle$ requires that we first obtain $\Phi^{m}_{+}(v,\theta)$ via Eq.~\eqref{Phimplus}. 
We evaluate the integral by taking advantage of the fact that both $e^{im\Omega_{\rm H} v'}$ and $\Psi^m(v',\theta)$ vary over a time scale ${\cal {R}}$ that is much longer than $\kappa^{-1} \sim 4M$. Integration by parts returns 
\begin{equation} 
\Phi^m_+ = \frac{1}{\kappa} e^{i m \Omega_{\rm H} v} \Psi^m \,, 
\end{equation} 
up to terms that are smaller by a factor of ${\cal{O}}(M/{\cal{R}}) = {\cal{O}}(\chi)$. Inserting this in Eq.~\eqref{Adot-def}, discarding corrections of ${\cal{O}}(\chi^2)$ and defining the new invariants
\begin{align}
E_6 &= \dot{\cal E}_{ab} \dot{\cal E}^{ab}\,, \qquad
B_6 = \dot{\cal B}_{ab} \dot{\cal B}^{ab}\,,
\end{align}
gives
\begin{align} 
\frac{\kappa}{8\pi} \langle\dot{A}\rangle &= \frac{16}{45} M^6 \Bigl[
4 \Omega_{\rm H}^2 \langle E_1 + B_1\rangle 
- 6 \Omega_{\rm H}^2 \langle E_2 + B_2 \rangle \nn \\
&- 4 \Omega_{\rm H} \langle E_4 +B_4 \rangle 
+ \langle E_6 + B_6 \rangle \Bigr]\,. 
\label{Adot_smallchi} 
\end{align}
In the limit $\chi \to 0$, this becomes 
\begin{equation} 
\frac{\kappa}{8\pi} \langle\dot{A}\rangle = \frac{16}{45} M^6 \langle E_6 + B_6 \rangle\,,
\end{equation} 
which agrees with the Schwarzschild expression in~\cite{Poisson:2004cw}. When $\Omega_{\rm H} \simeq \chi/(4M)$ is large compared with
${\cal R}^{-1}$, we obtain instead 
\begin{equation} 
\frac{\kappa}{8\pi} \langle\dot{A}\rangle \simeq \frac{32}{45} M^6 \Omega_{\rm H}^2  
\Bigl[ 2 \langle E_1 + B_1\rangle
- 3 \langle E_2 + B_2 \rangle\Bigr]\,, 
\end{equation} 
which agrees with Eq.~\eqref{Adotffinal} in the regime $\chi \ll 1$.  

In the case of rigid rotation, the tidal fields satisfy the relations given in Eq.~\eqref{rigid-relations} and through a similar calculation we can find that for the new invariant $E_6$, 
\be
E_6 = 2\Omega^2 ( 2E_1 - 3 E_2 )\,. 
\ee
A similar identity is true for $B_6$.
With these, it follows that Eq.~\eqref{Adot_smallchi} reduces to 
\begin{equation}\label{Adot_smallchirigid}
\frac{\kappa}{8\pi} \langle\dot{A}\rangle = \frac{32}{45} M^6 
\bigl( \Omega - \Omega_{\rm H} \bigr)^2 
\Bigl[ 2 \langle E_1 + B_1\rangle
- 3 \langle E_2 + B_2\rangle \Bigr] \,,
\end{equation} 
when the tidal fields are rigidly rotating. 

For rigid rotation, $\langle\dot{A}\rangle$ gives rise to separate expressions
for $\langle\dot{M}\rangle$ and $\langle\dot{J}\rangle$. We rely on the first law of BH mechanics, $(\kappa/8\pi)
\langle\dot{A}\rangle = \langle\dot{M}\rangle- \Omega_{\rm H} \langle\dot{J}\rangle$, and the rigid-rotation 
relation $\langle\dot{M}\rangle = \Omega \langle\dot{J}\rangle$, to obtain
\begin{equation} \label{Mdot_smallchirigid}
\langle\dot{M}\rangle =  \frac{32}{45} M^6 \Omega 
\bigl( \Omega - \Omega_{\rm H} \bigr) 
\Bigl[ 2 \langle E_1 + B_1\rangle
- 3 \langle E_2 + B_2\rangle \Bigr]\,, 
\end{equation} 
and 
\begin{equation} \label{Jdot_smallchirigid}
\langle\dot{J}\rangle =  \frac{32}{45} M^6 \bigl( \Omega - \Omega_{\rm H} \bigr)  
\Bigl[ 2 \langle E_1 + B_1\rangle
- 3 \langle E_2 + B_2\rangle \Bigr]\,.  
\end{equation} 
These expressions agree with the results in Sec.~IX F of
~\cite{Poisson:2004cw} in the regime $\chi \ll 1$. According to our
conventions, the sign of $\Omega$ is measured relative to the
BH rotation. Assuming that the tidal invariant within square
brackets is positive, we have that the BH gains mass when
$\Omega > 0$ and $\Omega > \Omega_{\rm H}$; otherwise the BH
loses mass (the tidal dynamics is superradiant). We also have that the
BH gains angular momentum when $\Omega > \Omega_{\rm H}$, and
loses angular momentum when $\Omega < \Omega_{\rm H}$. In all
circumstances, the tidal dynamics produce an increase in surface area.   

%%%%%%%%%%%%%%%%%%%%%%%%%%%%%%%%%%%%%%%%%%%%%%%%%%%%%%%%%%
\section{Applications: BH in a circular binary system in the slow-motion approximation}
\label{binary}

Let us apply the main results of Sec.~\ref{Psi-flux} to the case where the perturbed BH is in a circular binary around another BH with mass $M_{\ext}$. We consider the case where the orbital velocity is small (slow-motion approximation), $ V \ll 1$. If the BH is in a circular orbit with radius $b$, the orbital angular velocity and the relative orbital velocity are given by~\cite{Taylor:2008xy}
\begin{align} \label{Omegaslow}
\Omega &= \epsilon \sqrt{\frac{M_T}{b^3}}\left[ 1 - \frac{1}{2}(3 + \eta) V^2+{\cal{O}}(V^{4}) \right ] \,,\\ 
V &= \sqrt{\frac{M_T}{b}}\,,
\end{align}
respectively, where $\epsilon \equiv \hat{\mathbf{L}} \cdot \hat{\mathbf{s}} =\pm 1$, $\eta=M M_{\ext}/M_T^2$ is the symmetric mass ratio and $M_T=M+M_{\ext}$ the total mass. Then, the non vanishing frame components of the tidal tensors are given by~\cite{Taylor:2008xy}
\begin{align}
\frac{1}{2}(&{\cal{E}}_{11} + {\cal{E}}_{22}) = - \frac{M_{\ext}}{2b^3}\left[1 + \frac{M}{2M_T}V^2 + {\cal{O}}(V^4)\right] \label{Eslow-beg}\,, \\
\frac{1}{2}\!({\cal{E}}_{11} &\!-\! {\cal{E}}_{22}) \!=\! - \frac{3M_{\ext}}{2b^3}\!\left[1 \!-\! \frac{M_{\ext} \!+\! 3M_T}{2M_T}V^2 \!+\! {\cal{O}}(V^4)\right]\! \cos{2 \Omega t}, \\
{\cal{E}}_{12} &= - \frac{3M_{\ext}}{2b^3}\left[1 - \frac{M_{\ext} + 3M_T}{2M_T}V^2 + {\cal{O}}(V^4)\right] \sin{2 \Omega t} \,, \\
{\cal{B}}_{13} &=  -\frac{3 M_{\ext}}{b^3} V \cos{\Omega t} + {\cal{O}}(V^3)\,,\\
{\cal{B}}_{23} &=  -\frac{3 M_{\ext}}{b^3} V \sin{\Omega t} + {\cal{O}}(V^3)\label{Bslow-end}\,,
\end{align}
to the appropriate order.

Substituting Eqs.~\eqref{Eslow-beg}-\eqref{Bslow-end} into Eqs.~\eqref{Mdotffinal} and \eqref{Jdotffinal}, we find 
\begin{align}\label{Mdotslow}
\langle \dot{M} \rangle &= -\frac{8}{5} \epsilon \, \eta^2 \left(\frac{ M}{M_T}\right)^3 \chi (1 + 3 \chi^2) V^{15} \nn \\
&+ \frac{2}{5} \epsilon \, \eta^2 \left(\frac{ M}{M_T}\right)^3 \chi V^{17} \left[3(23 \chi^2 + 6) \right.\nn\\
&\left.- 2\left(1 + 3\chi^2\right) \frac{ M}{M_T} - 2\left(1 + 3\chi^2\right)\left(\frac{ M}{M_T}\right)^2 \right] \nn \\
&+\frac{16}{5}  \, \eta^2 \left(\frac{ M}{M_T}\right)^4\! (\sigma\!+\!1) (1 \!-\! 15 \chi^2 \!+\! 2\chi^4) V^{18} \!+\!{\cal{O}}(V^{19})\,,
\end{align}
and
\begin{align}\label{Jdotslow}
\langle \dot{J} \rangle &= -\frac{8}{5} \eta^2 \left(\frac{ M}{M_T}\right)^3 M_T \, \chi (1 + 3 \chi^2) V^{12} \nn \\
&+ \!\frac{2}{5} \eta^2 \!\left(\frac{ M}{M_T}\right)^3 \!M_T\! \left[(12 \!+\! 51 \chi^2) \!-\! (4 \!+\! 12 \chi^2)\frac{ M}{M_T}\right] \chi V^{14} \nn \\
&+\frac{16}{5} \epsilon \eta^2 \left(\frac{ M}{M_T}\right)^4M_T(\sigma+1)(1 - 15 \chi^2 + 2\chi^4) V^{15}\nn\\
&+ {\cal{O}}(V^{16})\,.
\end{align}
The third term in Eq.~\eqref{Mdotslow} is of higher order than what we are allowed to keep here. For circular orbits, however, $\langle \dot{M} \rangle= \Omega \langle \dot{J} \rangle$, and, thus, we can obtain this term from the third term in Eq.~\eqref{Jdotslow}. The results agree with~\cite{Poisson:2004cw} to ${\cal{O}}({\cal{R}}^{-2})$. Using the first law of BH mechanics, we find that the change in the horizon area is
\begin{align}\label{Adotslow}
\langle \dot{A} \rangle &=\frac{64 \pi}{5} \eta^2 \left(\frac{ M}{M_T}\right)^3 M_T \, \frac{\chi^2 (1 + 3 \chi^2)}{\sigma} V^{12} \nn \\
&- \!\frac{16 \pi}{5}\!\eta^2 \!\left(\!\frac{ M}{M_T}\!\right)^3\!\! M_T \!\left[(12 \!+\! 51 \chi^2) \!-\! (4 \!+\! 12 \chi^2)\frac{ M}{M_T}\right]\!\frac{\chi^2}{\sigma}\! V^{14}\nn \\
&-\frac{256 \pi}{5} \epsilon \eta^2 \left(\frac{ M}{M_T}\right)^4 M_T\frac{(\sigma + 1) (1-6\chi^2+\chi^4)\chi}{\sigma} V^{15}\nn\\
&+ {\cal{O}}(V^{16})\,.
\end{align}
As expected from the discussion of Sec.~\ref{rigid}, the fluxes are not subject to the $\alpha$ ambiguity. This is a result of the fact that the BH is in rigid rotation around its companion.

Finally, these expressions, kept to ${\cal{O}}(\chi^2)$, are also valid in the slow-rotation limit, as one can easily verify from direct substitution of the tidal tensors given in Eqs.~\eqref{Eslow-beg}-\eqref{Bslow-end} in Eqs.~\eqref{Adot_smallchirigid}-\eqref{Jdot_smallchirigid}.

%%%%%%%%%%%%%%%%%%%%%%%%%%%%%%%%%%%%%%%%%%%%%%%%%%%%%%%%%%%
\section{Conclusions}
\label{conclusions}
%%%%%%%%%%%%%%%%%%%%%%%%%%%%%%%%%%%%%%%%%%%%%%%%%%%%%%%%%%%

We studied how the properties of a rotating BH are modified when it is
subject to external vacuum perturbations, extending previous
calculations to ${\cal{O}}({\cal{R}}^{-3})$ (where ${\cal{R}}$ is the radius of curvature of the external geometry) in the slow-motion/small-hole
approximation. We mainly used the NP scalar $\psi_{0}$ to describe
these perturbations, where $\psi_{0}$ must satisfy the Teukolsky equation.    

The asymptotic behavior of $\psi_0$ depends on the external universe and it is constructed as follows. Working under the assumption that the external universe induces a small perturbation on the background spacetime and it varies slowly (that is, we expand all quantities in powers of $M/{\cal{R}}$, where $M$ is the mass of the BH), we parametrized the Weyl tensor of the external spacetime through time-dependent symmetric and trace-free tidal tensors ${\cal{E}}_{ab}$, ${\cal{B}}_{ab}$, ${\cal{E}}_{abc}$ and ${\cal{B}}_{abc}$. Then, the asymptotic form of $\psi_0$ can be obtained by Taylor expanding the Weyl tensor to the appropriate order in ${\cal{R}}^{-1}$ and expressed in terms of the above mentioned tidal tensors. Then, projecting the Weyl tensor onto a tetrad, one obtains the asymptotic form of $\psi_0$. 

This asymptotic NP scalar suggests a decomposition of the Fourier-transformed $\tilde{\psi}_0$ into multipole modes. The decomposition introduces some unknown radial functions that can be determined by solving the Teukolsky equation perturbatively in $\omega = {\cal{O}}({\cal{R}}^{-1}) \ll 1$. The resulting $\tilde{\psi}_{0}$ is then valid close to the unperturbed horizon, outside the asymptotic region. The time dependence of this full scalar is still given by the tidal tensors, while its angular dependence is given by spin-weighted spheroidal harmonics. 
To obtain the fluxes of energy and spin angular momentum across the horizon, we evaluated
$\psi_0$ on the (unperturbed) horizon with the Hartle-Hawking tetrad to construct the Teukolsky potential. 

No external universe structure is specified when obtaining these results. The only restriction imposed on the external universe is that it be slowly varying and its tidal effects on the background BH be small. Once the external universe is specified, for example, through a post-Newtonian metric, the tidal tensors could also be determined by asymptotic matching~\cite{Yunes:2005nn,Yunes:2006iw,JohnsonMcDaniel:2009dq}.

To conclude, we applied our results to an astrophysically motivated scenario: a binary system composed of two rotating BHs.  We assumed that the external universe perturbations are because of a Kerr BH and calculated the fluxes of mass and angular momentum and the change of horizon surface of the background BH for the case of a slow-motion orbit. 

The calculation of the mass and angular momentum fluxes across the BH horizon is a definite step toward understanding the perturbed dynamics around rotating BHs. These fluxes are necessary to self-consistently calculate the evolution of the orbit, and thus, the emitted gravitational radiation. However, to fully describe the perturbed spacetime, we need its metric, which could be a possible avenue for future research. The perturbed metric close to the event horizon of the perturbed BH can be constructed from the NP scalar through the Chrzanowski procedure~\cite{chrz,Ori:2002uv,Wald:1978vm,ck2,Yunes:2005ve}, and it would allow us to study the geometry of the perturbed horizon~\cite{Poisson:2009qj,Vega:2011ue}.  

By asymptotically matching this Kerr perturbed metric to a spinning binary PN metric that is valid far from the background BH, we can also obtain a full metric that describes the dynamics of the entire spacetime~\cite{Yunes:2005nn,Yunes:2006iw,JohnsonMcDaniel:2009dq,Gallouin:2012kb}. Such a metric could provide initial data for numerical relativity simulations~\cite{2012PhDT.........1C,Reifenberger:2012yg}, or it could serve as a background metric for accretion disk studies~\cite{Noble:2012xz}. 

%%%%%%%%%%%%%%%%%%%%%%%%%%%%%
\acknowledgments

We thank Luis Lehner for helpful discussions while working on this paper.  N.Y. acknowledges support from NSF Grant No. PHY-1114374 and NASA Grant No. NNX11AI49G, under 00001944. E.P. acknowledges support from the Natural Sciences and Engineering Research Council of Canada.

%%%%%%%%%%%%%%%%%%%%%%%%%%%%%%%%%%%%%%%%%%%%%%%
% APPENDICES
%%%%%%%%%%%%%%%%%%%%%%%%%%%%%%%%%%%%%%%%%%%%%%%

%%%%%%%%%%%%%%%%%%%%%%%%%%%%%%%%%%%%%%%%%%%%%%%
% REFERENCES
%%%%%%%%%%%%%%%%%%%%%%%%%%%%%%%%%%%%%%%%%%%%%%%
\bibliography{master.bib}

\begin{thebibliography}{44}%
\makeatletter
\providecommand \@ifxundefined [1]{%
 \@ifx{#1\undefined}
}%
\providecommand \@ifnum [1]{%
 \ifnum #1\expandafter \@firstoftwo
 \else \expandafter \@secondoftwo
 \fi
}%
\providecommand \@ifx [1]{%
 \ifx #1\expandafter \@firstoftwo
 \else \expandafter \@secondoftwo
 \fi
}%
\providecommand \natexlab [1]{#1}%
\providecommand \enquote  [1]{``#1''}%
\providecommand \bibnamefont  [1]{#1}%
\providecommand \bibfnamefont [1]{#1}%
\providecommand \citenamefont [1]{#1}%
\providecommand \href@noop [0]{\@secondoftwo}%
\providecommand \href [0]{\begingroup \@sanitize@url \@href}%
\providecommand \@href[1]{\@@startlink{#1}\@@href}%
\providecommand \@@href[1]{\endgroup#1\@@endlink}%
\providecommand \@sanitize@url [0]{\catcode `\\12\catcode `\$12\catcode
  `\&12\catcode `\#12\catcode `\^12\catcode `\_12\catcode `\%12\relax}%
\providecommand \@@startlink[1]{}%
\providecommand \@@endlink[0]{}%
\providecommand \url  [0]{\begingroup\@sanitize@url \@url }%
\providecommand \@url [1]{\endgroup\@href {#1}{\urlprefix }}%
\providecommand \urlprefix  [0]{URL }%
\providecommand \Eprint [0]{\href }%
\providecommand \doibase [0]{http://dx.doi.org/}%
\providecommand \selectlanguage [0]{\@gobble}%
\providecommand \bibinfo  [0]{\@secondoftwo}%
\providecommand \bibfield  [0]{\@secondoftwo}%
\providecommand \translation [1]{[#1]}%
\providecommand \BibitemOpen [0]{}%
\providecommand \bibitemStop [0]{}%
\providecommand \bibitemNoStop [0]{.\EOS\space}%
\providecommand \EOS [0]{\spacefactor3000\relax}%
\providecommand \BibitemShut  [1]{\csname bibitem#1\endcsname}%
\let\auto@bib@innerbib\@empty
%</preamble>
\bibitem [{\citenamefont {Kerr}(1963)}]{Kerr:1963ud}%
  \BibitemOpen
  \bibfield  {author} {\bibinfo {author} {\bibfnamefont {R.~P.}\ \bibnamefont
  {Kerr}},\ }\href {\doibase 10.1103/PhysRevLett.11.237} {\bibfield  {journal}
  {\bibinfo  {journal} {Phys.Rev.Lett.}\ }\textbf {\bibinfo {volume} {11}},\
  \bibinfo {pages} {237} (\bibinfo {year} {1963})}\BibitemShut {NoStop}%
%%CITATION = PRLTA,11,237;%%
\bibitem [{\citenamefont {Hawking}(1971)}]{Hawking:1971bv}%
  \BibitemOpen
  \bibfield  {author} {\bibinfo {author} {\bibfnamefont {S.}~\bibnamefont
  {Hawking}},\ }\href {\doibase 10.1007/BF00759218} {\bibfield  {journal}
  {\bibinfo  {journal} {Gen.Rel.Grav.}\ }\textbf {\bibinfo {volume} {1}},\
  \bibinfo {pages} {393} (\bibinfo {year} {1971})}\BibitemShut {NoStop}%
%%CITATION = GRGVA,1,393;%%
\bibitem [{\citenamefont {Hawking}(1972)}]{Hawking:1971vc}%
  \BibitemOpen
  \bibfield  {author} {\bibinfo {author} {\bibfnamefont {S.}~\bibnamefont
  {Hawking}},\ }\href {\doibase 10.1007/BF01877517} {\bibfield  {journal}
  {\bibinfo  {journal} {Commun.Math.Phys.}\ }\textbf {\bibinfo {volume} {25}},\
  \bibinfo {pages} {152} (\bibinfo {year} {1972})}\BibitemShut {NoStop}%
%%CITATION = CMPHA,25,152;%%
\bibitem [{\citenamefont {Israel}(1967)}]{PhysRev.164.1776}%
  \BibitemOpen
  \bibfield  {author} {\bibinfo {author} {\bibfnamefont {W.}~\bibnamefont
  {Israel}},\ }\href {\doibase 10.1103/PhysRev.164.1776} {\bibfield  {journal}
  {\bibinfo  {journal} {Phys. Rev.}\ }\textbf {\bibinfo {volume} {164}},\
  \bibinfo {pages} {1776} (\bibinfo {year} {1967})}\BibitemShut {NoStop}%
\bibitem [{\citenamefont {Israel}(1968)}]{Israel:1967za}%
  \BibitemOpen
  \bibfield  {author} {\bibinfo {author} {\bibfnamefont {W.}~\bibnamefont
  {Israel}},\ }\href@noop {} {\bibfield  {journal} {\bibinfo  {journal}
  {Commun.Math.Phys.}\ }\textbf {\bibinfo {volume} {8}},\ \bibinfo {pages}
  {245} (\bibinfo {year} {1968})}\BibitemShut {NoStop}%
%%CITATION = CMPHA,8,245;%%
\bibitem [{\citenamefont {Carter}(1971)}]{Carter:1971zc}%
  \BibitemOpen
  \bibfield  {author} {\bibinfo {author} {\bibfnamefont {B.}~\bibnamefont
  {Carter}},\ }\href {\doibase 10.1103/PhysRevLett.26.331} {\bibfield
  {journal} {\bibinfo  {journal} {Phys.Rev.Lett.}\ }\textbf {\bibinfo {volume}
  {26}},\ \bibinfo {pages} {331} (\bibinfo {year} {1971})}\BibitemShut
  {NoStop}%
%%CITATION = PRLTA,26,331;%%
\bibitem [{\citenamefont {Poisson}\ and\ \citenamefont
  {Sasaki}(1995)}]{Poisson:1994yf}%
  \BibitemOpen
  \bibfield  {author} {\bibinfo {author} {\bibfnamefont {E.}~\bibnamefont
  {Poisson}}\ and\ \bibinfo {author} {\bibfnamefont {M.}~\bibnamefont
  {Sasaki}},\ }\href {\doibase 10.1103/PhysRevD.51.5753} {\bibfield  {journal}
  {\bibinfo  {journal} {Phys.Rev.}\ }\textbf {\bibinfo {volume} {D51}},\
  \bibinfo {pages} {5753} (\bibinfo {year} {1995})},\ \Eprint
  {http://arxiv.org/abs/gr-qc/9412027} {arXiv:gr-qc/9412027 [gr-qc]}
  \BibitemShut {NoStop}%
%%CITATION = GR-QC/9412027;%%
\bibitem [{\citenamefont {Alvi}(2001)}]{Alvi:2001mx}%
  \BibitemOpen
  \bibfield  {author} {\bibinfo {author} {\bibfnamefont {K.}~\bibnamefont
  {Alvi}},\ }\href {\doibase 10.1103/PhysRevD.64.104020} {\bibfield  {journal}
  {\bibinfo  {journal} {Phys.Rev.}\ }\textbf {\bibinfo {volume} {D64}},\
  \bibinfo {pages} {104020} (\bibinfo {year} {2001})},\ \Eprint
  {http://arxiv.org/abs/gr-qc/0107080} {arXiv:gr-qc/0107080 [gr-qc]}
  \BibitemShut {NoStop}%
%%CITATION = GR-QC/0107080;%%
\bibitem [{\citenamefont {Tagoshi}\ \emph {et~al.}(1997)\citenamefont
  {Tagoshi}, \citenamefont {Mano},\ and\ \citenamefont
  {Takasugi}}]{Tagoshi:1997jy}%
  \BibitemOpen
  \bibfield  {author} {\bibinfo {author} {\bibfnamefont {H.}~\bibnamefont
  {Tagoshi}}, \bibinfo {author} {\bibfnamefont {S.}~\bibnamefont {Mano}}, \
  and\ \bibinfo {author} {\bibfnamefont {E.}~\bibnamefont {Takasugi}},\ }\href
  {\doibase 10.1143/PTP.98.829} {\bibfield  {journal} {\bibinfo  {journal}
  {Prog.Theor.Phys.}\ }\textbf {\bibinfo {volume} {98}},\ \bibinfo {pages}
  {829} (\bibinfo {year} {1997})},\ \Eprint
  {http://arxiv.org/abs/gr-qc/9711072} {arXiv:gr-qc/9711072 [gr-qc]}
  \BibitemShut {NoStop}%
%%CITATION = GR-QC/9711072;%%
\bibitem [{\citenamefont {Fang}\ and\ \citenamefont
  {Lovelace}(2005)}]{PhysRevD.72.124016}%
  \BibitemOpen
  \bibfield  {author} {\bibinfo {author} {\bibfnamefont {H.}~\bibnamefont
  {Fang}}\ and\ \bibinfo {author} {\bibfnamefont {G.}~\bibnamefont
  {Lovelace}},\ }\href {\doibase 10.1103/PhysRevD.72.124016} {\bibfield
  {journal} {\bibinfo  {journal} {Phys. Rev. D}\ }\textbf {\bibinfo {volume}
  {72}},\ \bibinfo {pages} {124016} (\bibinfo {year} {2005})}\BibitemShut
  {NoStop}%
\bibitem [{\citenamefont {Hughes}(2001)}]{Hughes:2001jr}%
  \BibitemOpen
  \bibfield  {author} {\bibinfo {author} {\bibfnamefont {S.~A.}\ \bibnamefont
  {Hughes}},\ }\href {\doibase 10.1103/PhysRevD.64.064004} {\bibfield
  {journal} {\bibinfo  {journal} {Phys.Rev.}\ }\textbf {\bibinfo {volume}
  {D64}},\ \bibinfo {pages} {064004} (\bibinfo {year} {2001})},\ \Eprint
  {http://arxiv.org/abs/gr-qc/0104041} {arXiv:gr-qc/0104041 [gr-qc]}
  \BibitemShut {NoStop}%
%%CITATION = GR-QC/0104041;%%
\bibitem [{\citenamefont {Martel}(2004)}]{Martel:2003jj}%
  \BibitemOpen
  \bibfield  {author} {\bibinfo {author} {\bibfnamefont {K.}~\bibnamefont
  {Martel}},\ }\href@noop {} {\bibfield  {journal} {\bibinfo  {journal} {Phys.
  Rev.}\ }\textbf {\bibinfo {volume} {D69}},\ \bibinfo {pages} {044025}
  (\bibinfo {year} {2004})},\ \Eprint {http://arxiv.org/abs/gr-qc/0311017}
  {gr-qc/0311017} \BibitemShut {NoStop}%
%%CITATION = GR-QC 0311017;%%
\bibitem [{\citenamefont {Price}\ and\ \citenamefont
  {Whelan}(2001)}]{Price:2001un}%
  \BibitemOpen
  \bibfield  {author} {\bibinfo {author} {\bibfnamefont {R.~H.}\ \bibnamefont
  {Price}}\ and\ \bibinfo {author} {\bibfnamefont {J.~T.}\ \bibnamefont
  {Whelan}},\ }\href {\doibase 10.1103/PhysRevLett.87.231101} {\bibfield
  {journal} {\bibinfo  {journal} {Phys.Rev.Lett.}\ }\textbf {\bibinfo {volume}
  {87}},\ \bibinfo {pages} {231101} (\bibinfo {year} {2001})},\ \Eprint
  {http://arxiv.org/abs/gr-qc/0107029} {arXiv:gr-qc/0107029 [gr-qc]}
  \BibitemShut {NoStop}%
%%CITATION = GR-QC/0107029;%%
\bibitem [{\citenamefont {Pound}\ and\ \citenamefont
  {Poisson}(2008)}]{Pound:2007th}%
  \BibitemOpen
  \bibfield  {author} {\bibinfo {author} {\bibfnamefont {A.}~\bibnamefont
  {Pound}}\ and\ \bibinfo {author} {\bibfnamefont {E.}~\bibnamefont
  {Poisson}},\ }\href {\doibase 10.1103/PhysRevD.77.044013} {\bibfield
  {journal} {\bibinfo  {journal} {Phys.Rev.}\ }\textbf {\bibinfo {volume}
  {D77}},\ \bibinfo {pages} {044013} (\bibinfo {year} {2008})},\ \Eprint
  {http://arxiv.org/abs/0708.3033} {arXiv:0708.3033 [gr-qc]} \BibitemShut
  {NoStop}%
%%CITATION = ARXIV:0708.3033;%%
\bibitem [{\citenamefont {Yunes}\ \emph {et~al.}(2010)\citenamefont {Yunes},
  \citenamefont {Buonanno}, \citenamefont {Hughes}, \citenamefont
  {Coleman~Miller},\ and\ \citenamefont {Pan}}]{Yunes:2009ef}%
  \BibitemOpen
  \bibfield  {author} {\bibinfo {author} {\bibfnamefont {N.}~\bibnamefont
  {Yunes}}, \bibinfo {author} {\bibfnamefont {A.}~\bibnamefont {Buonanno}},
  \bibinfo {author} {\bibfnamefont {S.~A.}\ \bibnamefont {Hughes}}, \bibinfo
  {author} {\bibfnamefont {M.}~\bibnamefont {Coleman~Miller}}, \ and\ \bibinfo
  {author} {\bibfnamefont {Y.}~\bibnamefont {Pan}},\ }\href {\doibase
  10.1103/PhysRevLett.104.091102} {\bibfield  {journal} {\bibinfo  {journal}
  {Phys.Rev.Lett.}\ }\textbf {\bibinfo {volume} {104}},\ \bibinfo {pages}
  {091102} (\bibinfo {year} {2010})},\ \Eprint {http://arxiv.org/abs/0909.4263}
  {arXiv:0909.4263 [gr-qc]} \BibitemShut {NoStop}%
%%CITATION = ARXIV:0909.4263;%%
\bibitem [{\citenamefont {Yunes}\ \emph {et~al.}(2011)\citenamefont {Yunes},
  \citenamefont {Buonanno}, \citenamefont {Hughes}, \citenamefont {Pan},
  \citenamefont {Barausse}, \citenamefont {Miller},\ and\ \citenamefont
  {Throwe}}]{PhysRevD.83.044044}%
  \BibitemOpen
  \bibfield  {author} {\bibinfo {author} {\bibfnamefont {N.}~\bibnamefont
  {Yunes}}, \bibinfo {author} {\bibfnamefont {A.}~\bibnamefont {Buonanno}},
  \bibinfo {author} {\bibfnamefont {S.~A.}\ \bibnamefont {Hughes}}, \bibinfo
  {author} {\bibfnamefont {Y.}~\bibnamefont {Pan}}, \bibinfo {author}
  {\bibfnamefont {E.}~\bibnamefont {Barausse}}, \bibinfo {author}
  {\bibfnamefont {M.~C.}\ \bibnamefont {Miller}}, \ and\ \bibinfo {author}
  {\bibfnamefont {W.}~\bibnamefont {Throwe}},\ }\href {\doibase
  10.1103/PhysRevD.83.044044} {\bibfield  {journal} {\bibinfo  {journal} {Phys.
  Rev. D}\ }\textbf {\bibinfo {volume} {83}},\ \bibinfo {pages} {044044}
  (\bibinfo {year} {2011})}\BibitemShut {NoStop}%
\bibitem [{\citenamefont {Poisson}(2004)}]{Poisson:2004cw}%
  \BibitemOpen
  \bibfield  {author} {\bibinfo {author} {\bibfnamefont {E.}~\bibnamefont
  {Poisson}},\ }\href {\doibase 10.1103/PhysRevD.70.084044} {\bibfield
  {journal} {\bibinfo  {journal} {Phys.Rev.}\ }\textbf {\bibinfo {volume}
  {D70}},\ \bibinfo {pages} {084044} (\bibinfo {year} {2004})},\ \Eprint
  {http://arxiv.org/abs/gr-qc/0407050} {arXiv:gr-qc/0407050 [gr-qc]}
  \BibitemShut {NoStop}%
%%CITATION = GR-QC/0407050;%%
\bibitem [{\citenamefont {Regge}\ and\ \citenamefont
  {{Wheeler}}(1957)}]{Regge:1957rw}%
  \BibitemOpen
  \bibfield  {author} {\bibinfo {author} {\bibfnamefont {T.}~\bibnamefont
  {Regge}}\ and\ \bibinfo {author} {\bibfnamefont {J.~A.}\ \bibnamefont
  {{Wheeler}}},\ }\href {\doibase 10.1103/PhysRev.108.1063} {\bibfield
  {journal} {\bibinfo  {journal} {Phys. Rev.}\ }\textbf {\bibinfo {volume}
  {108}},\ \bibinfo {pages} {1063} (\bibinfo {year} {1957})}\BibitemShut
  {NoStop}%
\bibitem [{\citenamefont {Zerilli}(1970)}]{Zerilli:1970la}%
  \BibitemOpen
  \bibfield  {author} {\bibinfo {author} {\bibfnamefont {F.~J.}\ \bibnamefont
  {Zerilli}},\ }\href@noop {} {\bibfield  {journal} {\bibinfo  {journal}
  {\prd}\ }\textbf {\bibinfo {volume} {2}},\ \bibinfo {pages} {2141} (\bibinfo
  {year} {1970})}\BibitemShut {NoStop}%
\bibitem [{\citenamefont {Teukolsky}(1973)}]{teukolsky}%
  \BibitemOpen
  \bibfield  {author} {\bibinfo {author} {\bibfnamefont {S.~A.}\ \bibnamefont
  {Teukolsky}},\ }\href {\doibase 10.1086/152444} {\bibfield  {journal}
  {\bibinfo  {journal} {Astrophys. J.}\ }\textbf {\bibinfo {volume} {185}},\
  \bibinfo {pages} {635} (\bibinfo {year} {1973})}\BibitemShut {NoStop}%
%%CITATION = ASJOA,185,635;%%
\bibitem [{\citenamefont {Press}\ and\ \citenamefont
  {Teukolsky}(1973)}]{Press:1973zz}%
  \BibitemOpen
  \bibfield  {author} {\bibinfo {author} {\bibfnamefont {W.~H.}\ \bibnamefont
  {Press}}\ and\ \bibinfo {author} {\bibfnamefont {S.~A.}\ \bibnamefont
  {Teukolsky}},\ }\href {\doibase 10.1086/152445} {\bibfield  {journal}
  {\bibinfo  {journal} {Astrophys.J.}\ }\textbf {\bibinfo {volume} {185}},\
  \bibinfo {pages} {649} (\bibinfo {year} {1973})}\BibitemShut {NoStop}%
%%CITATION = ASJOA,185,649;%%
\bibitem [{\citenamefont {Teukolsky}\ and\ \citenamefont
  {Press}(1974)}]{Teukolsky:1974yv}%
  \BibitemOpen
  \bibfield  {author} {\bibinfo {author} {\bibfnamefont {S.}~\bibnamefont
  {Teukolsky}}\ and\ \bibinfo {author} {\bibfnamefont {W.}~\bibnamefont
  {Press}},\ }\href {\doibase 10.1086/153180} {\bibfield  {journal} {\bibinfo
  {journal} {Astrophys.J.}\ }\textbf {\bibinfo {volume} {193}},\ \bibinfo
  {pages} {443} (\bibinfo {year} {1974})}\BibitemShut {NoStop}%
%%CITATION = ASJOA,193,443;%%
\bibitem [{\citenamefont {Poisson}(2005)}]{Poisson:2005pi}%
  \BibitemOpen
  \bibfield  {author} {\bibinfo {author} {\bibfnamefont {E.}~\bibnamefont
  {Poisson}},\ }\href {\doibase 10.1103/PhysRevLett.94.161103} {\bibfield
  {journal} {\bibinfo  {journal} {Phys.Rev.Lett.}\ }\textbf {\bibinfo {volume}
  {94}},\ \bibinfo {pages} {161103} (\bibinfo {year} {2005})},\ \Eprint
  {http://arxiv.org/abs/gr-qc/0501032} {arXiv:gr-qc/0501032 [gr-qc]}
  \BibitemShut {NoStop}%
\bibitem [{\citenamefont {Martel}\ and\ \citenamefont
  {Poisson}(2005)}]{Martel:2005ir}%
  \BibitemOpen
  \bibfield  {author} {\bibinfo {author} {\bibfnamefont {K.}~\bibnamefont
  {Martel}}\ and\ \bibinfo {author} {\bibfnamefont {E.}~\bibnamefont
  {Poisson}},\ }\href@noop {} {\bibfield  {journal} {\bibinfo  {journal} {Phys.
  Rev.}\ }\textbf {\bibinfo {volume} {D71}},\ \bibinfo {pages} {104003}
  (\bibinfo {year} {2005})},\ \Eprint {http://arxiv.org/abs/gr-qc/0502028}
  {gr-qc/0502028} \BibitemShut {NoStop}%
%%CITATION = GR-QC 0502028;%%
\bibitem [{\citenamefont {Taylor}\ and\ \citenamefont
  {Poisson}(2008)}]{Taylor:2008xy}%
  \BibitemOpen
  \bibfield  {author} {\bibinfo {author} {\bibfnamefont {S.}~\bibnamefont
  {Taylor}}\ and\ \bibinfo {author} {\bibfnamefont {E.}~\bibnamefont
  {Poisson}},\ }\href {\doibase 10.1103/PhysRevD.78.084016} {\bibfield
  {journal} {\bibinfo  {journal} {Phys.Rev.}\ }\textbf {\bibinfo {volume}
  {D78}},\ \bibinfo {pages} {084016} (\bibinfo {year} {2008})},\ \Eprint
  {http://arxiv.org/abs/0806.3052} {arXiv:0806.3052 [gr-qc]} \BibitemShut
  {NoStop}%
%%CITATION = ARXIV:0806.3052;%%
\bibitem [{\citenamefont {Comeau}\ and\ \citenamefont
  {Poisson}(2009)}]{Comeau:2009bz}%
  \BibitemOpen
  \bibfield  {author} {\bibinfo {author} {\bibfnamefont {S.}~\bibnamefont
  {Comeau}}\ and\ \bibinfo {author} {\bibfnamefont {E.}~\bibnamefont
  {Poisson}},\ }\href {\doibase 10.1103/PhysRevD.80.087501} {\bibfield
  {journal} {\bibinfo  {journal} {Phys.Rev.}\ }\textbf {\bibinfo {volume}
  {D80}},\ \bibinfo {pages} {087501} (\bibinfo {year} {2009})},\ \Eprint
  {http://arxiv.org/abs/0908.4518} {arXiv:0908.4518 [gr-qc]} \BibitemShut
  {NoStop}%
%%CITATION = ARXIV:0908.4518;%%
\bibitem [{\citenamefont {Poisson}\ and\ \citenamefont
  {Vlasov}(2010)}]{Poisson:2009qj}%
  \BibitemOpen
  \bibfield  {author} {\bibinfo {author} {\bibfnamefont {E.}~\bibnamefont
  {Poisson}}\ and\ \bibinfo {author} {\bibfnamefont {I.}~\bibnamefont
  {Vlasov}},\ }\href {\doibase 10.1103/PhysRevD.81.024029} {\bibfield
  {journal} {\bibinfo  {journal} {Phys.Rev.}\ }\textbf {\bibinfo {volume}
  {D81}},\ \bibinfo {pages} {024029} (\bibinfo {year} {2010})},\ \Eprint
  {http://arxiv.org/abs/0910.4311} {arXiv:0910.4311 [gr-qc]} \BibitemShut
  {NoStop}%
%%CITATION = ARXIV:0910.4311;%%
\bibitem [{\citenamefont {Vega}\ \emph {et~al.}(2011)\citenamefont {Vega},
  \citenamefont {Poisson},\ and\ \citenamefont {Massey}}]{Vega:2011ue}%
  \BibitemOpen
  \bibfield  {author} {\bibinfo {author} {\bibfnamefont {I.}~\bibnamefont
  {Vega}}, \bibinfo {author} {\bibfnamefont {E.}~\bibnamefont {Poisson}}, \
  and\ \bibinfo {author} {\bibfnamefont {R.}~\bibnamefont {Massey}},\ }\href
  {\doibase 10.1088/0264-9381/28/17/175006} {\bibfield  {journal} {\bibinfo
  {journal} {Class.Quant.Grav.}\ }\textbf {\bibinfo {volume} {28}},\ \bibinfo
  {pages} {175006} (\bibinfo {year} {2011})},\ \Eprint
  {http://arxiv.org/abs/1106.0510} {arXiv:1106.0510 [gr-qc]} \BibitemShut
  {NoStop}%
%%CITATION = ARXIV:1106.0510;%%
\bibitem [{\citenamefont {Yunes}\ and\ \citenamefont
  {Gonzalez}(2006)}]{Yunes:2005ve}%
  \BibitemOpen
  \bibfield  {author} {\bibinfo {author} {\bibfnamefont {N.}~\bibnamefont
  {Yunes}}\ and\ \bibinfo {author} {\bibfnamefont {J.~A.}\ \bibnamefont
  {Gonzalez}},\ }\href@noop {} {\bibfield  {journal} {\bibinfo  {journal}
  {Phys. Rev.}\ }\textbf {\bibinfo {volume} {D73}},\ \bibinfo {pages} {024010}
  (\bibinfo {year} {2006})},\ \Eprint {http://arxiv.org/abs/gr-qc/0510076}
  {gr-qc/0510076} \BibitemShut {NoStop}%
%%CITATION = GR-QC 0510076;%%
\bibitem [{\citenamefont {{Teukolsky}}(1972)}]{Teukolsky:1972le}%
  \BibitemOpen
  \bibfield  {author} {\bibinfo {author} {\bibfnamefont {S.~A.}\ \bibnamefont
  {{Teukolsky}}},\ }\href {\doibase 10.1103/PhysRevLett.29.1114} {\bibfield
  {journal} {\bibinfo  {journal} {\prl}\ }\textbf {\bibinfo {volume} {29}},\
  \bibinfo {pages} {1114} (\bibinfo {year} {1972})}\BibitemShut {NoStop}%
\bibitem [{\citenamefont {Mano}\ \emph {et~al.}(1996)\citenamefont {Mano},
  \citenamefont {Suzuki},\ and\ \citenamefont {Takasugi}}]{Mano:1996vt}%
  \BibitemOpen
  \bibfield  {author} {\bibinfo {author} {\bibfnamefont {S.}~\bibnamefont
  {Mano}}, \bibinfo {author} {\bibfnamefont {H.}~\bibnamefont {Suzuki}}, \ and\
  \bibinfo {author} {\bibfnamefont {E.}~\bibnamefont {Takasugi}},\ }\href
  {\doibase 10.1143/PTP.95.1079} {\bibfield  {journal} {\bibinfo  {journal}
  {Prog.Theor.Phys.}\ }\textbf {\bibinfo {volume} {95}},\ \bibinfo {pages}
  {1079} (\bibinfo {year} {1996})},\ \Eprint
  {http://arxiv.org/abs/gr-qc/9603020} {arXiv:gr-qc/9603020 [gr-qc]}
  \BibitemShut {NoStop}%
%%CITATION = GR-QC/9603020;%%
\bibitem [{\citenamefont {Mino}\ \emph {et~al.}(1997)\citenamefont {Mino},
  \citenamefont {Sasaki},\ and\ \citenamefont {Tanaka}}]{Mino:1997bw}%
  \BibitemOpen
  \bibfield  {author} {\bibinfo {author} {\bibfnamefont {Y.}~\bibnamefont
  {Mino}}, \bibinfo {author} {\bibfnamefont {M.}~\bibnamefont {Sasaki}}, \ and\
  \bibinfo {author} {\bibfnamefont {T.}~\bibnamefont {Tanaka}},\ }\href
  {\doibase 10.1143/PTPS.128.373} {\bibfield  {journal} {\bibinfo  {journal}
  {Prog.Theor.Phys.Suppl.}\ }\textbf {\bibinfo {volume} {128}},\ \bibinfo
  {pages} {373} (\bibinfo {year} {1997})},\ \Eprint
  {http://arxiv.org/abs/gr-qc/9712056} {arXiv:gr-qc/9712056 [gr-qc]}
  \BibitemShut {NoStop}%
%%CITATION = GR-QC/9712056;%%
\bibitem [{\citenamefont {Alvi}(2000)}]{Alvi:1999cw}%
  \BibitemOpen
  \bibfield  {author} {\bibinfo {author} {\bibfnamefont {K.}~\bibnamefont
  {Alvi}},\ }\href {\doibase 10.1103/PhysRevD.61.124013} {\bibfield  {journal}
  {\bibinfo  {journal} {Phys. Rev.}\ }\textbf {\bibinfo {volume} {D61}},\
  \bibinfo {pages} {124013} (\bibinfo {year} {2000})},\ \Eprint
  {http://arxiv.org/abs/gr-qc/9912113} {arXiv:gr-qc/9912113} \BibitemShut
  {NoStop}%
%%CITATION = GR-QC/9912113;%%
\bibitem [{\citenamefont {Yunes}\ \emph {et~al.}(2005)\citenamefont {Yunes},
  \citenamefont {Tichy}, \citenamefont {Owen},\ and\ \citenamefont
  {Bruegmann}}]{Yunes:2005nn}%
  \BibitemOpen
  \bibfield  {author} {\bibinfo {author} {\bibfnamefont {N.}~\bibnamefont
  {Yunes}}, \bibinfo {author} {\bibfnamefont {W.}~\bibnamefont {Tichy}},
  \bibinfo {author} {\bibfnamefont {B.~J.}\ \bibnamefont {Owen}}, \ and\
  \bibinfo {author} {\bibfnamefont {B.}~\bibnamefont {Bruegmann}},\ }\href@noop
  {} {\  (\bibinfo {year} {2005})},\ \Eprint
  {http://arxiv.org/abs/gr-qc/0503011} {gr-qc/0503011} \BibitemShut {NoStop}%
%%CITATION = GR-QC 0503011;%%
\bibitem [{\citenamefont {Yunes}\ and\ \citenamefont
  {Tichy}(2006)}]{Yunes:2006iw}%
  \BibitemOpen
  \bibfield  {author} {\bibinfo {author} {\bibfnamefont {N.}~\bibnamefont
  {Yunes}}\ and\ \bibinfo {author} {\bibfnamefont {W.}~\bibnamefont {Tichy}},\
  }\href {\doibase 10.1103/PhysRevD.74.064013} {\bibfield  {journal} {\bibinfo
  {journal} {Phys. Rev.}\ }\textbf {\bibinfo {volume} {D74}},\ \bibinfo {pages}
  {064013} (\bibinfo {year} {2006})},\ \Eprint
  {http://arxiv.org/abs/gr-qc/0601046} {arXiv:gr-qc/0601046} \BibitemShut
  {NoStop}%
%%CITATION = GR-QC/0601046;%%
\bibitem [{\citenamefont {Gallouin}\ \emph {et~al.}(2012)\citenamefont
  {Gallouin}, \citenamefont {Nakano}, \citenamefont {Yunes},\ and\
  \citenamefont {Campanelli}}]{Gallouin:2012kb}%
  \BibitemOpen
  \bibfield  {author} {\bibinfo {author} {\bibfnamefont {L.}~\bibnamefont
  {Gallouin}}, \bibinfo {author} {\bibfnamefont {H.}~\bibnamefont {Nakano}},
  \bibinfo {author} {\bibfnamefont {N.}~\bibnamefont {Yunes}}, \ and\ \bibinfo
  {author} {\bibfnamefont {M.}~\bibnamefont {Campanelli}},\ }\href@noop {} {\
  (\bibinfo {year} {2012})},\ \Eprint {http://arxiv.org/abs/1208.6489}
  {arXiv:1208.6489 [gr-qc]} \BibitemShut {NoStop}%
%%CITATION = ARXIV:1208.6489;%%
\bibitem [{\citenamefont {Johnson-McDaniel}\ \emph {et~al.}(2009)\citenamefont
  {Johnson-McDaniel}, \citenamefont {Yunes}, \citenamefont {Tichy},\ and\
  \citenamefont {Owen}}]{JohnsonMcDaniel:2009dq}%
  \BibitemOpen
  \bibfield  {author} {\bibinfo {author} {\bibfnamefont {N.~K.}\ \bibnamefont
  {Johnson-McDaniel}}, \bibinfo {author} {\bibfnamefont {N.}~\bibnamefont
  {Yunes}}, \bibinfo {author} {\bibfnamefont {W.}~\bibnamefont {Tichy}}, \ and\
  \bibinfo {author} {\bibfnamefont {B.~J.}\ \bibnamefont {Owen}},\ }\href
  {\doibase 10.1103/PhysRevD.80.124039} {\bibfield  {journal} {\bibinfo
  {journal} {Phys.Rev.}\ }\textbf {\bibinfo {volume} {D80}},\ \bibinfo {pages}
  {124039} (\bibinfo {year} {2009})},\ \Eprint {http://arxiv.org/abs/0907.0891}
  {arXiv:0907.0891 [gr-qc]} \BibitemShut {NoStop}%
\bibitem [{\citenamefont {Chrzanowski}(1975)}]{chrz}%
  \BibitemOpen
  \bibfield  {author} {\bibinfo {author} {\bibfnamefont {P.~L.}\ \bibnamefont
  {Chrzanowski}},\ }\href {\doibase 10.1103/PhysRevD.11.2042} {\bibfield
  {journal} {\bibinfo  {journal} {Phys. Rev.}\ }\textbf {\bibinfo {volume}
  {D11}},\ \bibinfo {pages} {2042} (\bibinfo {year} {1975})}\BibitemShut
  {NoStop}%
%%CITATION = PHRVA,D11,2042;%%
\bibitem [{\citenamefont {Ori}(2003)}]{Ori:2002uv}%
  \BibitemOpen
  \bibfield  {author} {\bibinfo {author} {\bibfnamefont {A.}~\bibnamefont
  {Ori}},\ }\href {\doibase 10.1103/PhysRevD.67.124010} {\bibfield  {journal}
  {\bibinfo  {journal} {Phys.Rev.}\ }\textbf {\bibinfo {volume} {D67}},\
  \bibinfo {pages} {124010} (\bibinfo {year} {2003})},\ \Eprint
  {http://arxiv.org/abs/gr-qc/0207045} {arXiv:gr-qc/0207045 [gr-qc]}
  \BibitemShut {NoStop}%
%%CITATION = GR-QC/0207045;%%
\bibitem [{\citenamefont {Wald}(1978)}]{Wald:1978vm}%
  \BibitemOpen
  \bibfield  {author} {\bibinfo {author} {\bibfnamefont {R.~M.}\ \bibnamefont
  {Wald}},\ }\href {\doibase 10.1103/PhysRevLett.41.203} {\bibfield  {journal}
  {\bibinfo  {journal} {Phys.Rev.Lett.}\ }\textbf {\bibinfo {volume} {41}},\
  \bibinfo {pages} {203} (\bibinfo {year} {1978})}\BibitemShut {NoStop}%
%%CITATION = PRLTA,41,203;%%
\bibitem [{\citenamefont {Kegeles}\ and\ \citenamefont {Cohen}(1979)}]{ck2}%
  \BibitemOpen
  \bibfield  {author} {\bibinfo {author} {\bibfnamefont {L.~S.}\ \bibnamefont
  {Kegeles}}\ and\ \bibinfo {author} {\bibfnamefont {J.~M.}\ \bibnamefont
  {Cohen}},\ }\href {\doibase 10.1103/PhysRevD.19.1641} {\bibfield  {journal}
  {\bibinfo  {journal} {Phys. Rev.}\ }\textbf {\bibinfo {volume} {D19}},\
  \bibinfo {pages} {1641} (\bibinfo {year} {1979})}\BibitemShut {NoStop}%
%%CITATION = PHRVA,D19,1641;%%
\bibitem [{\citenamefont {{Chu}}(2012)}]{2012PhDT.........1C}%
  \BibitemOpen
  \bibfield  {author} {\bibinfo {author} {\bibfnamefont {T.}~\bibnamefont
  {{Chu}}},\ }\emph {\bibinfo {title} {{Numerical simulations of black-hole
  spacetimes}}},\ \href@noop {} {Ph.D. thesis},\ \bibinfo  {school} {California
  Institute of Technology} (\bibinfo {year} {2012})\BibitemShut {NoStop}%
\bibitem [{\citenamefont {Reifenberger}\ and\ \citenamefont
  {Tichy}(2012)}]{Reifenberger:2012yg}%
  \BibitemOpen
  \bibfield  {author} {\bibinfo {author} {\bibfnamefont {G.}~\bibnamefont
  {Reifenberger}}\ and\ \bibinfo {author} {\bibfnamefont {W.}~\bibnamefont
  {Tichy}},\ }\href {\doibase 10.1103/PhysRevD.86.064003} {\bibfield  {journal}
  {\bibinfo  {journal} {Phys.Rev.}\ }\textbf {\bibinfo {volume} {D86}},\
  \bibinfo {pages} {064003} (\bibinfo {year} {2012})},\ \Eprint
  {http://arxiv.org/abs/1205.5502} {arXiv:1205.5502 [gr-qc]} \BibitemShut
  {NoStop}%
%%CITATION = ARXIV:1205.5502;%%
\bibitem [{\citenamefont {Noble}\ \emph {et~al.}(2012)\citenamefont {Noble},
  \citenamefont {Mundim}, \citenamefont {Nakano}, \citenamefont {Krolik},
  \citenamefont {Campanelli} \emph {et~al.}}]{Noble:2012xz}%
  \BibitemOpen
  \bibfield  {author} {\bibinfo {author} {\bibfnamefont {S.~C.}\ \bibnamefont
  {Noble}}, \bibinfo {author} {\bibfnamefont {B.~C.}\ \bibnamefont {Mundim}},
  \bibinfo {author} {\bibfnamefont {H.}~\bibnamefont {Nakano}}, \bibinfo
  {author} {\bibfnamefont {J.~H.}\ \bibnamefont {Krolik}}, \bibinfo {author}
  {\bibfnamefont {M.}~\bibnamefont {Campanelli}},  \emph {et~al.},\ }\href
  {\doibase 10.1088/0004-637X/755/1/51} {\bibfield  {journal} {\bibinfo
  {journal} {Astrophys.J.}\ }\textbf {\bibinfo {volume} {755}},\ \bibinfo
  {pages} {51} (\bibinfo {year} {2012})},\ \Eprint
  {http://arxiv.org/abs/1204.1073} {arXiv:1204.1073 [astro-ph.HE]} \BibitemShut
  {NoStop}%
%%CITATION = ARXIV:1204.1073;%%
\end{thebibliography}%

\end{document}